\documentclass[twocolumn]{aastex631}

\newcommand{\cms}{\ensuremath{\rm cm\,s^{-1}}}
\newcommand{\ms}{\ensuremath{\rm m\,s^{-1}}}
\providecommand{\me}{\ensuremath{\,M_{\rm E}}}

\definecolor{my_color}{HTML}{3a18b1}

\definecolor{new_color}{HTML}{CF0000}

\definecolor{html_black}{HTML}{000000}

\definecolor{new_color}{RGB}{189,39,7}

\begin{document}

\title{Identifying Exoplanets with Deep Learning VI. Enhancing neural network mitigation of stellar activity RV signals with additional metrics}

\author[0009-0004-4014-0937]{Naomi McWilliam}
\affiliation{Department of Physics, Imperial College London, Prince Consort Road, London, SW7 2AZ, UK}
\affiliation{Department of Physics and Kavli Institute for Astrophysics and Space Research, Massachusetts Institute of Technology, Cambridge, MA 02139, USA}

\author[0000-0002-7564-6047]{Zo{\"e}\ L. de Beurs}
\affiliation{Department of Earth, Atmospheric and Planetary Sciences, Massachusetts Institute of Technology, Cambridge, MA 02139, USA}
\thanks{NSF Graduate Research Fellow, MIT Presidential Fellow, MIT Collamore-Rogers Fellow}

\author[0000-0001-7246-5438]{Andrew Vanderburg}
\affiliation{Department of Physics and Kavli Institute for Astrophysics and Space Research, Massachusetts Institute of Technology, Cambridge, MA 02139, USA}

\author[0000-0002-0563-784X]{Javier Via\~{n}a}
\affiliation{Department of Physics and Kavli Institute for Astrophysics and Space Research, Massachusetts Institute of Technology, Cambridge, MA 02139, USA}

\author[0000-0001-7254-4363]{Annelies Mortier}
\affiliation{School of Physics \& Astronomy, University of Birmingham, Edgbaston, Birmingham, B15 2TT, UK}

\author[0000-0003-1605-5666]{Lars A. Buchhave}
\affiliation{DTU Space,  Technical University of Denmark, Elektrovej 328, DK-2800 Kgs. Lyngby, Denmark}

\author[0000-0002-8863-7828]{Andrew Collier Cameron}
\affiliation{Center for Exoplanet Science, SUPA, School of Physics \& Astronomy
University of St Andrews, North Haugh St Andrews, Fife, KY16 9SS}

\author[0000-0003-1784-1431]{Rosario Cosentino}
\affiliation{Fundación Galileo Galilei—INAF, Rambla J.A. F. Perez, 7, E-38712 S.C. Tenerife, Spain}

\author[0000-0002-9332-2011]{Xavier Dumusque}
\affiliation{Observatoire de Genève, Université de Genèvee, 51 chemin des Maillettes, 1290 Versoix, Switzerland}

\author[0000-0003-4702-5152]{Adriano Ghedina}
\affiliation{Fundación Galileo Galilei—INAF, Rambla J.A. F. Perez, 7, E-38712 S.C. Tenerife, Spain}

\author[0000-0002-8122-2240]{Ben Lakeland}
\affiliation{School of Physics \& Astronomy, University of Birmingham, Edgbaston, Birmingham, B15 2TT, UK}
\affiliation{Department of Physics and Astronomy, University of Exeter, Exeter, EX4 4QL, UK}

\author[0000-0002-0651-4294]{Marcello Lodi}
\affiliation{Fundación Galileo Galilei—INAF, Rambla J.A. F. Perez, 7, E-38712 S.C. Tenerife, Spain}

\author[0000-0003-3204-8183]{Mercedes L\'opez-Morales}
\affiliation{Space Telescope Science Institute, 3700 San Martin Drive, Baltimore MD 21218, USA}

\author[0000-0001-7014-1771]{Dimitar Sasselov}
\affiliation{Center for Astrophysics | Harvard \& Smithsonian, 60 Garden St, Cambridge, MA 02138, USA}

\author[0000-0001-7014-1771]{Alessandro Sozzetti}
\affiliation{INAF-Osservatorio Astrofisico di Torino, via Osservatorio 20, I-10025 Pino Torinese, Italy}



\begin{abstract}

The measurement of exoplanet masses using the radial velocity (RV) technique is currently limited by stellar activity, which introduces quasiperiodic variability signals that must be modeled and removed to enhance the sensitivity of the RV measurements to exoplanet signals. Neural networks have previously been demonstrated effective in modeling stellar activity signals in HARPS-N solar data using white light cross correlation functions (CCFs). Building on this work, we train a neural network on six years of HARPS-N solar data with additional parameters commonly associated to stellar activity, including chromatic CCFs, line shape metrics, spectral activity indicators, total solar irradiance (TSI) light curves from SORCE and TSIS-1, and TSI time derivatives. Our results show that parameters such as the bisector inverse slope and Na D equivalent widths do not significantly improve the neural network's ability to predict activity-induced RV variations compared to using the white light CCFs alone. However, parameters such as unsigned magnetic flux, the TSI and its time derivative, S-index, H-alpha equivalent width, chromatic CCFs, contrast, and full width at half maximum do improve the neural network's ability to predict RV scatter. Our new model reduces the RV scatter in a held-out test set from 147.1 \cms\ to 93.3 \cms, consistent with supergranulation noise levels reported in previous studies. These results suggest that finding effective tracers for (super)granulation will be critical to train models capable of further mitigating RV jitter, and necessary for characterizing Earth analogues.

\end{abstract}


\section{Introduction} \label{sec:intro}
The radial velocity (RV) method is one of the most widely used tools for exoplanet detection and enables the detection of one of the most fundamental properties of an exoplanet: its mass
\footnote{It is important to note the RV method measures $m\sin i$, where $m$ is the mass of the exoplanet and $i$ is the inclination of its orbit relative to the line of sight. Therefore, the RV method alone cannot determine the true mass of the exoplanet without additional information about the orbital inclination, which must be obtained from other observational methods.}
\citep{Fischer_2016}. The RV method measures line-of-sight velocities through the observed Doppler shifts in the spectra of stars.
However, RV measurements have currently hit a noise floor of $\sim$50-100 \cms\ that poses a barrier to achieving the extreme precision radial velocities (EPRVs) necessary for the detection of Earth analogue exoplanets. This noise originates from stellar surface phenomena and is often referred to as stellar activity or stellar jitter \citep[e.g.][]{Lovis2010, Fischer_2016, 2020miklosHaywood}. Characterizing and effectively modeling stellar variability signals is critical to detecting and characterizing potentially Earth-like exoplanets orbiting Sun-like stars.

Stellar activity signals evolve in time, resulting in quasi-periodic signals that can be challenging to model \citep[e.g. ][]{haywood2016}. The stellar activity signals that primarily limit RV precision for Sun-like stars are caused by four physical processes: (i) pressure-mode oscillations (p-modes) on the stellar surface \citep{1962leighton} which cause RV signals of 10 to 100 \cms\ for solar-like stars \citep{2000Schrijver, 2008arentoft} on timescales of a few minutes for the Sun \citep{2000Schrijver, 2009broomhall},(ii) granulation \citep{1982dravins, 1982livingston, 1990brandt, 2011Dumusque} and supergranulation \citep{Rieutord2010, Bazot2012} phenomena, which produce signals with similar amplitudes to p-modes \citep{2000Schrijver} on timescales from a few minutes up to a couple days in the Sun \citep{1989title, 2004DelMoro, AlMoulla2023},  (iii) spots and faculae \citep{1982livingston, 1982dravins, 1985Cavallini,2010Meunier, 1997saar, 2001queloz, Hu_lamo_2008, 2010lagrange} which can produce RV scatter on the order of $\sim 40 - 140$ \cms\ for the Sun \citep{2010Meunier} due to suppressed convection in active regions and result in signals on timescales of tens of days \citep{2011Dumusque}, and (iv) solar-like magnetic cycles on the timescale of several years \citep{1985dravins, 1988campbell,2003Lindegren,2010Meunier}.

Characterizing and removing stellar activity signals is particularly crucial given the current -- HARPS-N \citep{Cosentino2012HARPSN}, ESPRESSO \citep{2021A&A...645A..96P}, EXPRES \citep{2020AJ....159..187P}, NEID \citep{2016SPIE.9908E..7HS}, MAROON-X \citep{2022SPIE12184E..1GS}, KPF \citep{2020SPIE11447E..42G} -- and upcoming high-resolution spectrographs -- G-CLEF \citep{2014Szentgyorgyi}, HARPS-3 \citep{2016SPIE.9908E..6FT}, ANDES \citep{Marconi2024} --. These instruments already have \citep{2016Anglada, 2020suarez} or are expected to demonstrate instrumental RV precision required for the detection of Earth-mass exoplanets in the habitable zone around M-dwarfs (e.g. HARPS can achieve RV measurements of better than 100 \cms\ in short exposure time for bright stars \citep{2005pepe, 2011Dumusque}). However, accurate modeling of stellar variability will be critical to detecting Earth analogues around Sun-like stars which are expected to produce signals of about $10$ \cms.

Several methods have been developed to reduce the RV noise originating from stellar activity. For example, the HARPS-GTO survey uses exposure times of around $15$ minutes to reduce the effects from solar p-modes on RV measurements \citep{2011Dumusque}. \citet{2019chaplin} have shown that the p-modes can be averaged out to around $\sim 10$ \cms\ for Sun-like stars by fine-tuning exposure times, while \citet{2018Medina} were able to extend this by removing the p-mode effects from evolved stars. But longer exposure times are not an efficient way to reduce the variability from granulation or supergranulation phenomena, which act on timescales longer than single exposures \citep{2011Dumusque, Meunier2015}. Instead, \citet{2011Dumusque} showed that averaging multiple observations of a star separated by a few hours can significantly reduce these short-term effects, but not to the 10 \cms\ level needed to find Earth analogues.

The EPRV community has aimed much work at mitigating magnetic activity at either the rotation period or long-term magnetic cycle timescales. There are two main classes of solutions: those that involve separating activity signals from planetary signals in the time domain, and those that involve separating the signals in the spectral or wavelength domain.

In the time domain, early efforts used noise models like autoregressive moving averages \citep{Tuomi2013A&A} or perfectly periodic Keplerian signals \citep{Dumusque2012Natur}, but most work now takes advantage of Gaussian process regression \citep[e.g.][]{Haywood2014MNRAS, Rajpaul2015MNRAS, Rajpaul2016, Jones2017David}. However, these methods often rely on high-cadence and precisely timed observations, which can be difficult to achieve for astronomical observations.

In the spectral or wavelength domain, many mitigation efforts focus on identifying metrics that track activity signals and use those metrics as proxies to decorrelate activity from RV measurements. These methods involve indicators such as $\log R'_\mathrm{HK}$ \citep{1984Noyes}, the bisector inverse slope \citep{2001queloz}, H$\alpha$ \citep{Bonfils2007, Robertson2014}, and a combination of the star's light curve and its derivative (called $FF'$) \citep{Aigrain2012MNRAS}. Recently, \citet{Haywood_2022} used the unsigned magnetic flux to identify stellar activity, and \citet{Lienhard2023} demonstrated a method to extract a proxy of the unsigned magnetic flux in disk-integrated spectra and showed that it correlates strongly with RV variability. In the past few years, another class of solutions uses data driven-methods to separate stellar activity from true Doppler RV shifts in the spectral domain. An approach called YARARAv2 \citep{2023Cretignier} uses the RVs derived from individual lines to identify signals in the time series that can be filtered using Principle Component Analysis (PCA). This method has achieved sub \ms\ velocities on stars over timescales of decades.

One class of spectral domain-methods focuses on using cross-correlation functions (CCFs), which are obtained by cross-correlating observed spectra with a stellar mask, and encode both Doppler shifts and spectral line-shape variations. These methods use changes in the shape of the CCF to characterize and then mitigate stellar activity signals, and a range of techniques have been developed to model CCFs. Some of these techniques focus on using separating the shift-driven from shape-driven components of the CCF using PCA\citep[e.g.][]{CollierCameron2021, Klein2024} or Fourier domain methods \citep{Zhao2022}. One such approach, \texttt{SCALPELS}, uses PCA on the shift-invariant autocorrelation function of the CCF (which preserves CCF shape information only) and decorrelates against the strongest vectors to mitigate stellar activity \citep{CollierCameron2021, John2022}, yielding significant improvements in the RVs of active stars. 
Other techniques use forward modeling approaches, such as \citet{DiMaio2024} who developed \texttt{SpotCCF} to forward model CCF shape changes caused by starspots.

Beyond either time-domain or wavelength-domain methods, there has also been work to combine both of these approaches through the use of timeseries of the CCF that are modeled with a Gaussian process model \citep{Yu_Haochuan_2024}, which shows promise for using both spectral and timing information to model stellar variability.

In addition to these CCF-based methods that use Gaussian processes or PCA-based methods, \citet{deBeurs2022} have demonstrated that Machine learning (ML) methods such as neural networks (NN) can learn activity-driven variations in CCFs. These methods perform very well on both short-term (rotation) and long-term (magnetic cycle) activity. This is also a wavelength-domain method and therefore does not require precisely timed observations. \citet{deBeurs2022} demonstrated that white light CCFs contain shape changes that can be used as input to a NN capable of reducing stellar RV variability. Using this method, they reduced the RV variability in three years of observations of the Sun from the HARPS-N Solar Telescope \citep{Phillips2016} from 175.3 \cms\ to 103.9 \cms. Since then, several other ML techniques have been developed to mitigate this RV variability and demonstrated success on both simulated and real data \citep{2022Zhao, 2023Perger, Liang2023, 2023Colwell, deBeurs2024}.

In this paper, we build on the work by \citet{deBeurs2022}, by expanding their analysis of three years of HARPS-N Solar Telescope RVs to six years \footnote{This roughly doubles the volume of data used by \citet{deBeurs2022}}, and by testing several other activity indicators as inputs to their NN, in addition to the white light CCF used in that work. In ML, a feature is a measurable property or characteristic of a dataset that can help a NN identify and learn patterns. Often a range of data inputs are tested as features to determine which ones are most informative, as selecting the most informative features is crucial for optimizing the performance of ML models. Throughout this paper, we will refer to data inputs to our NN as features. We test features including the total solar irradiance (TSI), the time-derivative of TSI, unsigned magnetic flux, and the equivalent widths (EWs) for the H$\alpha$ and Sodium D1 and D2 absorption lines in the Sun's spectra. We also test additional features derived from the white light CCFs, such as the full-width half maximum (FWHM), the bisector span, and the contrast. Finally, we include chromatic CCFs measured from blue ($387.4-459.0$ nm), yellow ($453.9-554.0$ nm), and red ($547.9-690.9$ nm) wavelength regions of the solar spectra.

This paper is organized as follows. In Section \ref{sec:data}, we describe the observational data used for training the ML model. In Section \ref{sec:Methods}, we describe the methods used to process the input data in order to be suitable for the ML model. In Section \ref{sec: Neural Network Section}, we describe the mathematical foundations of our ML models, how the observations were separated into training and test sets, and the training procedure. In Section \ref{sec: Results}, we present our results. In Section \ref{sec: Discussion}, we discuss the implications of our results, and in Section \ref{sec: Conclusion} we conclude.

\section{Data} \label{sec:data}

\subsection{HARPS-N Solar Spectra} \label{sec:harps}
The RV data were obtained from the HARPS-N (High Accuracy Radial velocity Planet Searcher- North) optical spectrograph, which observes the Sun continuously with 5-minute integration times to mitigate short-term stellar activity $p$-mode oscillations \citep{Phillips2016, Dumusque2015}. HARPS-N is a temperature stabilized, cross-dispersed, $R \sim115,000$, echelle spectrograph, covering an optical wavelength range of $383-690 \ \mathrm{nm}$ \citep{Cosentino2012HARPSN}.

The solar data from the 5-minute exposures of the Sun, taken throughout the day using HARPS-N, are reduced using the HARPS-N Data Reduction Software (DRS-2.3.5) \citep{Dumusque2021}. In brief, the DRS extracts a 1D sky-background-subtracted spectrum for each echelle order and solves for the spectrograph's wavelength solution. The data undergoes cross-correlation with a digital mask based on solar absorption lines. This results in a 49-element array which we refer to as the cross-correlation function (CCF) for each echelle order. After corrections for instrumental drift are applied, the CCFs from each echelle order are summed to produce a white light high signal-to-noise ratio (SNR) CCF, which is used as the input representation for the ML method employed in the analysis described in Section \ref{sec: Neural Network Section}. The DRS then extracts the RVs by fitting the white light CCF with a Gaussian function. However, stellar variability including spots and faculae causes time-varying small shape deviations in the white light CCF from a perfect Gaussian, which result in the measured RVs including contributions from both Doppler shifts and stellar activity signals.

Unlike distant stars, the Sun is a resolved disk in the sky, so additional steps must be taken to correct for the fact that light from different regions of the Solar disk pass along different lines of sight through Earth's atmosphere. First, we account for the effect of differential atmospheric extinction following \citet{CollierCameron2019}. Then, we identify and remove observations where clouds obscured at least part of the solar disk following a multi-step process. First, we calculate a quality factor, based on the SNR and airmass of each observation, using a mixture model as described in \citet{CollierCameron2019}. We only include data where this quality factor (which spans 0 and 1) is above $0.99$. As this technique does not detect all low-quality observations, we additionally use information from the HARPS-N exposure meter to more sensitively identify observations contaminated by cloud coverage. We define the exposure meter quality factor $E$ as the ratio between the maximum and mean count of the exposure meter. This ratio does not pile up at 1 due to a slight delay in opening the shutter but can be well modeled by a Gaussian function to identify outlying observations where $E$ is large. After fitting the distribution of $E$, we remove all data where $E$ is higher than $3\sigma$ from the mean value of the entire data set. Finally, as a last cut, we also remove observations where the resulting RV is more than $5\sigma$ from the mean value. This rather strict approach of removing low-quality data may indeed remove good data too, but is purposefully strict to only have the highest quality of data. Although this process removes 35\% of the spectra, the number of days in the dataset is reduced by only 20\% since we take a daily average for each day. The dataset from HARPS-N after applying these quality cuts contains $1148$ days of solar observations between 2015-07-29 to 2021-11-12 
\footnote{The first three years of data from July 2015 - July 2018 can be downloaded publicly from \url{https://dace.unige.ch/sun/} (DRS version 2.2.8) or using the DACE python API \url{https://dace-query.readthedocs.io/en/latest/} (DRS version 2.3.5). The remaining RVs and activity indicators used in our analysis will be available via VizieR at CDS. \label{footnote: data}}.

We use multiple data products from the HARPS-N Solar observations in our analysis, which we describe in detail in Section \ref{sec:Methods}.

\subsection{Total Solar Irradiance Observations} \label{sec:tsi}
In addition to the HARPS-N solar spectra, we used observations of the total solar irradiance (TSI) of the Sun. The TSI is the intensity of solar radiation integrated over the solar surface, and over the entire spectrum of the Sun. The TSI can be thought of as an analogous data product to a light curve observed by \textit{Kepler} or \textit{TESS}, and therefore contains information about stellar activity \citep[e.g.][]{Aigrain2012MNRAS}. For example, increases in the TSI can be caused by the presence of faculae, while decreases can indicate the presence of sunspots. 

We downloaded TSI datasets from the Total Irradiance Monitor (TIM) instruments on the SORCE satellite \citep{SORCE-paper, Kopp2005, SORCE-data} and on the TSIS-1 sensor on the International Space Station \citep{TSIS-data}. The TIM on SORCE covers the entire solar spectrum -- from X-ray to far-infrared --, and has a high absolute accuracy of 350 parts per million (ppm), and relative accuracy of 10 ppm per year. It takes measurements at a 50-second cadence, which are combined to produce a daily averaged TSI value. The data set includes daily averaged TSI data from 2003-02-25 until 2020-02-25\footnote{This data can be downloaded publicly from \url{https://lasp.colorado.edu/sorce/data/tsi-data/}.}. On dates where the quality of data is low due to shielding of the Sun from the Earth, the TSI is recorded as 0.0. We removed these observations from the dataset.

The data from the TIM on the TSIS-1 sensor covers the same wavelength range, but with an improved absolute accuracy of 100 ppm, and relative accuracy of 10 ppm per year. \citep{TSIS-data}. The data set includes daily averaged TSI data from 2018-01-11 until 2023-07-19.\footnote{This data can be downloaded publicly from \url{https://lasp.colorado.edu/tsis/data/tsi-data/}.}

We cross-referenced the dates with measurements of TSI with HARPS-N Solar Telescope observations and found that all but 36 days with HARPS-N observations had a corresponding TSI measurement. We excluded those 36 days from our ML analysis to ensure that we had values for all input features included in our models.

\subsection{Helioseismic and Magnetic Imager Data} \label{sec: data - helioseismic and magnetic imager data}
To measure the disc-averaged, unsigned, unpolarized magnetic flux, we used data from the Helioseismic and Magnetic Imager \citep[HMI][]{Scherrer2012} aboard NASA's Solar Dynamics Observatory (SDO). The HMI observes the solar disc in the Fe I absorption line at $617.3\ \mathrm{nm}$. It achieves a high image resolution of $1$ arcsecond by observing the solar disc using two $4096\times4096$ pixel CCD cameras, and an image stabilization system, with a data rate of 55 Mbps. We used HMI's line of sight magnetogram data, which are maps of the Sun's photospheric magnetic field. The magnetograms have a precision of 10 G, a zero-point accuracy of 0.05 G, and are taken with a cadence of 45 seconds.

The line-of-sight magnetogram data were downloaded and processed using the open-source python package \texttt{SolAster} \footnote{\url{https://github.com/tamarervin/SolAster.git}} \citep{ervin-SolAster}, which we also used to calculate the unsigned magnetic flux from the magnetogram data, as described in Section \ref{sec: methods - unsigned magnetic flux}. \texttt{SolAster} downloads the data using the Sunpy \citep{sunpy_community2020} query software. For each date, three files are downloaded, a Dopplergram, a Magnetogram, and a Continuum Intensity. Each of these files correspond to the measurement at 12:00:45 UTC for that day. We downloaded daily observations for the full time period of the HARPS-N Solar observations. During the HARPS-N observational baseline, there were 9 dates where the HMI data were incomplete and not all three files were available, which meant the unsigned magnetic flux could not be calculated for those dates. We therefore excluded these dates and removed them from all of the datasets used in our analysis.

\section{Methods} \label{sec:Methods}

\subsection{Preparing CCFs and RVs}\label{sec:ccfs}
ML models learn most effectively when data is pre-processed and scaled in a uniform format to ensure each feature is of the same scale and considered equally \citep[e.g.][]{SINGH2020}. This pre-processing allows our ML models to capture and learn patterns in our data more readily. For the white light CCFs, we design the input representation such that the model becomes sensitive to shape changes and not translational shifts.\footnote{We note that our choice of preprocessing to highlight line shape changes means that our method will be less sensitive to activity signals that do not have significant line shape changes like granulation or supergranulation \citep{Meunier2020}.}
The measured RVs from the CCFs contain not only true velocity shifts from Doppler reflex motion but also contain stellar activity contributions and instrumental systematics. This is because the RVs from the DRS are computed by fitting a Gaussian to the white light CCFs. However, stellar activity in the form of spots and faculae, as well as instrumental systematics, can cause shape changes to CCFs that result in asymmetries. This means that the center measured by the Gaussian fit is not the true center (i.e. the true velocity shift), but instead the true velocity shift and some additional deviation. We want to train our model to predict this deviation between these center-of-the-CCF measurements and the true velocity shift. In this way, our model then aims to predict the contribution of CCF asymmetries to the RV measurements, and thereby models shape-driven stellar activity and instrumental noise, leaving us with shape-driven corrections for our RV timeseries.

\begin{figure}
    \centering
    \includegraphics[width=\columnwidth]{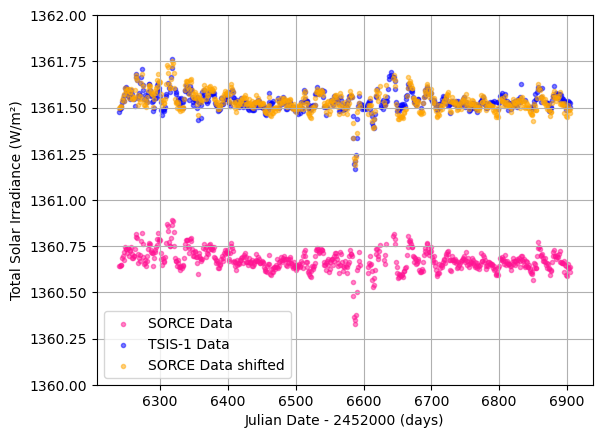}
    \caption{The overlapping region between the two data sets of TSI values. Pink points represent measurements from SORCE, while blue points represent measurements from TSIS-1. There is an offset between the two, so we shifted the SORCE data to match the level of the TSIS data to merge the datasets. The shifted SORCE data are shown in orange. Small variations can be seen between the TSI values from the two data sets.}
    \label{fig:tsi shift}
\end{figure}

The steps we take to design the input representation to predict differences between the Gaussian fit CCFs and the true velocity shifts are described in detail in \citet{deBeurs2022, deBeurs2024}, and summarized here:

\begin{enumerate}
    \item \textit{Weighted average CCFs.} We compute a SNR weighted average of the CCFs as described in detail in Section 2.1
    \item \textit{Remove the RV signatures of Solar System planets.} Since the RVs and CCFs from the DRS contain both the Doppler reflex motion of the Solar System planets and stellar variability, we transform both datatypes from the barycentric to the heliocentric reference frame using the JPL Horizons ephemeris \citep{Giorgini1996}. We tested multiple interpolation methods (linear, quadratic, cubic) and found that cubic was most optimal. After performing this transformation, the RVs and CCFs should contain only shifts due to stellar activity and instrumental systematics.
    \item \textit{Center the CCFs to the mean.} We fit a Gaussian to each CCF and shift all the CCFs to a common center using the expression $x_i = \rm mean(\mu) -\mu_i$, where $x_i$ is the amount shifted, $\mu_i$ is the center of the Gaussian fit for the CCF spectrum $i$, and $\rm mean (\mu)$ is the mean value of $\mu$ across the entire dataset.
    This step is taken because we know that planets cause translation shifts whereas stellar variability causes shape changes to the CCFs. By centering all the CCFs we remove the translational differences between the observations and help the model focus on learning shape change patterns instead. We note that centering the CCFs does not require knowing the planetary reflex motion \textit{a priori} and can be done for stars with unknown planetary contributions by using either pipeline-provided RVs or fitting a Gaussian and shifting by $\mu$.
    \item \textit{CCF normalization.} First, we normalize the CCFs by their continuum level. Then, to prepare the CCFs for the neural network, we subtract the median of the CCF timeseries and divide by the standard deviation. This normalization is performed across all input parameters as described below so that the scale of variations of each input parameter is approximately equal and speeds up training and optimization of the model.
\end{enumerate}

In addition to the standard white light CCF that uses all 69 spectral orders to compute a weighted average CCF, we generate three chromatic CCFs that use only a subset of the 69 orders: a) a blue CCF that is computed using the first 23 orders ($387.4-459.0$ nm), b) a yellow CCF that uses the middle 23 orders ($453.9-554.0$ nm), and c) a red CCF that uses the last 23 orders ($547.9-690.9$ nm). These chromatic CCFs are pre-processed in the same way as described above to ensure that they have the same normalized input representation. However, these chromatic CCFs each probe only a subset of spectral lines compared to the standard white light CCF and thus could encode activity information that is wavelength dependent. The presence of H-opacity, particularly in the blue wavelengths, allows us to probe deeper into the photosphere compared to red wavelengths. Therefore, variations in RVs across the chromatic CCFs could reveal differences in the hemisphere-averaged convective blueshift at different depths in the photosphere \citep{AlMoulla2022}.

\subsection{Sodium Doublet and H$\alpha$ Lines} \label{sec:Na}

In addition to the CCFs, we included the EWs of the H$\alpha$ line and the sodium D lines in the solar spectra as features to our ML model.
We computed the EWs of these lines in each HARPS-N spectrum using the python package \texttt{specutils.analysis.equivalent\_width} \citep{SpecUtils}, which uses as inputs the continuum normalized spectra, and the small wavelength regions centered around each absorption line summarized in Table \ref{table: EW wavelengths}. The code outputs EWs for each absorption line, calculated as the numerically integrated area between the observed spectrum and the normalized continuum level within the specified wavelength ranges.

\begin{deluxetable}{ccc}
\tablecaption{\label{table: EW wavelengths} Absorption lines and their corresponding wavelengths in air used to measure EWs. Lower and upper wavelength limits for each region used in the EW calculations are also listed.}
\tablehead{\colhead{Absorption Line} & \colhead{Wavelength in air} & \colhead{Wavelength cut region} \\ 
[-1.4ex]
\colhead{} & \colhead{(nm)} & \colhead{(nm)} } 
\startdata
H$\alpha$ &  656.281 &    589.508 - 589.685 \\
Na D1 &  589.592 &    589.508 - 589.685 \\
Na D2 &  588.995 &   588.879 - 589.105 \\
\enddata
\end{deluxetable}

We calculated the mean EW per day for each of the three absorption lines. We normalized the remaining EW values (subtracting the mean of the values and dividing by the standard deviation, similar to the CCF normalization), leaving us with three new features corresponding to the three absorption lines.

\subsection{TSI and TSI derivative}\label{sec: methods tsi}
We have two TSI data sets from TSIS-1 and SORCE which span different dates as described in Section \ref{sec:tsi}. These two data sets, containing daily averaged TSI values, overlap from 2018-01-11 until 2020-02-25, and there is an offset between them, with TSIS-1 having higher TSI values compared to SORCE as seen in Figure \ref{fig:tsi shift}. We also noticed that the first $100$ days of the TSIS-1 data had a significantly larger scatter compared to the SORCE data, which may be caused by initial instrument calibrations during the first period of operations for TSIS-1. 

To merge the datasets and remove the anomalous measurements at the beginning of the TSIS-1 time series, we took the steps described below. To start, we removed the first $110$ days of the TSIS-1 data for our analysis, which results in a smaller overlapping region between the two datasets that spans 2018-05-01 to 2020-02-25. We then computed the mean of the difference between the two data sets within this region and added this value to all the SORCE data as seen in Figure \ref{fig:tsi shift}. Then, for the overlapping region, we took the mean of the two daily averaged TSI values for those days. For dates where we only had values for one of the two datasets, those values were used. Lastly, we normalized all the TSI data in the same way as all the other input features (subtracting the mean and dividing by standard deviation) before feeding them into the neural network.

In addition to using the TSI values, we also computed the time derivative of the TSI, since this has also been shown to be important for predicting stellar activity signals, such as when using the FF' method \citep{Aigrain2012MNRAS}. In particular, sunspots, which move across the surface of the Sun, could potentially be identified through changes in TSI.

\begin{figure}
    \centering
    \includegraphics[width=\columnwidth]{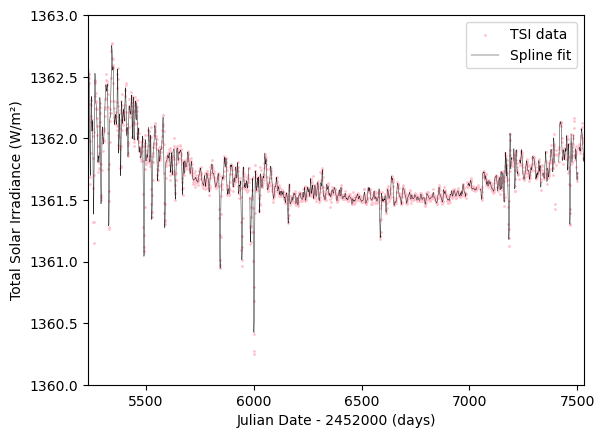}
    \caption{Plot of the total solar irradiance data from the merged SORCE and TSIS-1 datasets fitted with a spline using \texttt{keplerspline}. Pink points are individual TSI measurements, and the black curve is the spline fit. The spline captures the variations in the TSI values while smoothing over short-timescale variations. We use this spline to calculate the TSI derivative as described in Section \ref{sec: methods tsi}.}
    \label{fig:tsi spline}
\end{figure}

To find the change in the TSI, we first fit a spline to the un-normalized TSI data using \texttt{keplerspline} \footnote{\url{https://github.com/avanderburg/keplersplinev2}}, as shown in Figure \ref{fig:tsi spline}, where we used on average a $3.4$ day break point spacing. Then, we used the central difference scheme to numerically differentiate the spline. The derivative values were normalized, again by subtracting the median value and dividing by the standard deviation.

\subsection{Unsigned Magnetic Flux} \label{sec: methods - unsigned magnetic flux}

Another feature used in our model is the disc averaged, unsigned, unpolarized, magnetic flux of the Sun, which we obtain using the \texttt{SolAster} package, as described in \ref{sec: data - helioseismic and magnetic imager data}. The package identifies and removes regions without significant magnetic activity, where the unsigned magnetic field, $B_{obs}$, is less than a threshold chosen to be 8G. \texttt{SolAster} then calculates the disc averaged, unpolarized, unsigned magnetic flux, 
$|\hat{B_{obs}}|$, using
    \begin{equation}
        |\hat{B_{obs}}| = \frac{\sum_{\alpha\beta}|B_{obs,\alpha\beta}| I_{\alpha\beta}}{\sum_{\alpha\beta} I_{\alpha\beta}},
    \end{equation}
where $\alpha$ and $\beta$ are the coordinates on each point on the solar disc, $I_{\alpha\beta}$ is the uncorrected continuum intensity map, and where $B_{obs,\alpha\beta}$ is the corrected observed magnetic field strength from the magnetogram \citep{haywood2016,ervin-SolAster}.

After computing the unsigned magnetic flux using \texttt{SolAster}, we finally normalize this feature (again by subtracting the median and dividing by the standard deviation) such that it can be used as an NN input feature.

\subsection{S-Index}
We also used the S-Index (the ratio of the flux in the core of the Calcium II H and K lines to that in the continuum, e.g. \citealt{Vaughan1978PASP}) as a feature. These values were produced by the HARPS-N DRS \footref{footnote: data}. In the DRS, ghost contamination is automatically corrected \citep[see][]{Dumusque2021}. Before inputting into the neural network, the S-Index required some pre-processing. The raw data contained a significant number of outliers, which we identified using the same quality factor we used to identify good CCFs  \citep{CollierCameron2019}. We only kept data where the quality flag is over 0.99. This cut removed 42\% of the data points, but only 31\% of dates, most of which correspond to dates when the CCF data were also low quality. This additional cut therefore removed only 90 additional dates from the dataset. We then normalized the S-index measurements in the same way as all other input features.

\subsection{Final Data Set}

We only retained dates in our data set for training and evaluation of the neural network when there were measurements for all input features we considered. After removing all dates with either missing or outlier values from each independent metric, the final data set consisted of daily average observations on 978 individual dates between 2015-07-29 and 2021-09-08. 

\section{Neural Network Analysis} \label{sec: Neural Network Section}
We trained several Convolutional Neural Networks (CNNs) to predict stellar activity RV signals using the shape of the white light CCF and one or more different additional features.

\subsection{Neural Networks}
We use feedforward neural networks, where the network takes an input $\mathbf{x}$, and attempts to approximate a function $f^*$, so that it can output the value $y$, where $y=f(\mathbf{x};\mathbf{p})$, and $\mathbf{p}$ are the parameters it learns to produce the best approximation of $f^*$. In our case, $f^*$ is the true stellar activity correction and $f$ is our neural network approximation. The network is composed of many simpler functions. For example, the network might be the chain,
\begin{equation}
    f(\mathbf{x})=f^{(3)}(f^{(2)}(f^{(1)}(\mathbf{x}))),
\end{equation}
where $f^{(1)}$, $f^{(2)}$ and $f^{(3)}$ are three functions, and $f^{(1)}$ is the first layer, $f^{(2)}$ is the second layer and $f^{(3)}$ is the final output layer. The output from one layer is the input to the next layer. The length of the chain, in this case three, is called the depth of the model. We want $f(\mathbf{x})$ to match $f^*(\mathbf{x})$ as closely as possible.

To train the neural network, we used labelled training data that consists of input and output pairs $(\mathbf{x}, y)$. The neural network determines how to model the other functions in between the input and output, which are called hidden layers. The width of the model is given by the dimensions of the hidden layers. We can think of each layer as being composed of many units, where each layer takes many inputs from the previous layer, and produces a single output.

\subsection{Convolutional Neural Network Theory}
CNNs are commonly used for pattern recognition in structured data where the proximity of the dimensions contains relevant information. CNNs learn to identify local patterns across the entire input space. There are three types of layers in a CNN: convolutional layers, pooling layers, and fully connected layers. With each layer $i$, the complexity of the CNN increases, a larger portion of the input can be identified, and the final layer produces an output based on the entire input.

\subsubsection{Convolutional Layers}
We can first consider the convolutional layers, where we apply a cross-correlation operation. For each layer, $i$, we apply a 1-dimensional discrete CCF from a stack of $T$ vectors $\mathbf{a}_{i-1}^{(t)}$ for $t=1$ to $T$ of length $n_{i-1}$ called the input, to the stack of $L$ output vectors $\mathbf{a}_i^{(l)}$ for $l=1$ to $L$, or feature map, 
\begin{equation} \label{eq: CNN layer}
    \mathbf{a}_i^{(l)} = \phi \left( \sum_{t=1}^T \mathbf{w}_i^{(t,l)} * \mathbf{a}_{i-1}^{(t)} + \mathbf{b}_i^{(l)} \right),
\end{equation}
where the convolution kernel, $\mathbf{w}_i^{(t,l)}$, is a vector of length $m_i$ of learned parameters during training, $\mathbf{b}_i^{(l)}$ is a vector of length $n_i$ of learned bias parameters, and $\phi$ is an activation function as described in Section \ref{sec:activation functions}. The $*$ represents the convolution operator where for a vector $x$, feature map $s$, and kernel $w$,
\begin{equation}
    s_{i} = (w * x)_{i} = \sum_m x_{i+m} w_m.
\end{equation}
The kernel size is less than the input vector size so that it can be applied several times across regions of $a$. It is usually small, around $m_i=3$ or $5$, so that it can effectively detect local feature changes of $a$. Typically $m_i$ is odd as then the units from the previous layer can be centered around an output pixel, which would otherwise not be possible and cause distortion for an even valued $m_i$.

\subsubsection{Pooling Layers}
\begin{figure}
    \centering  \includegraphics[width=0.8\columnwidth]{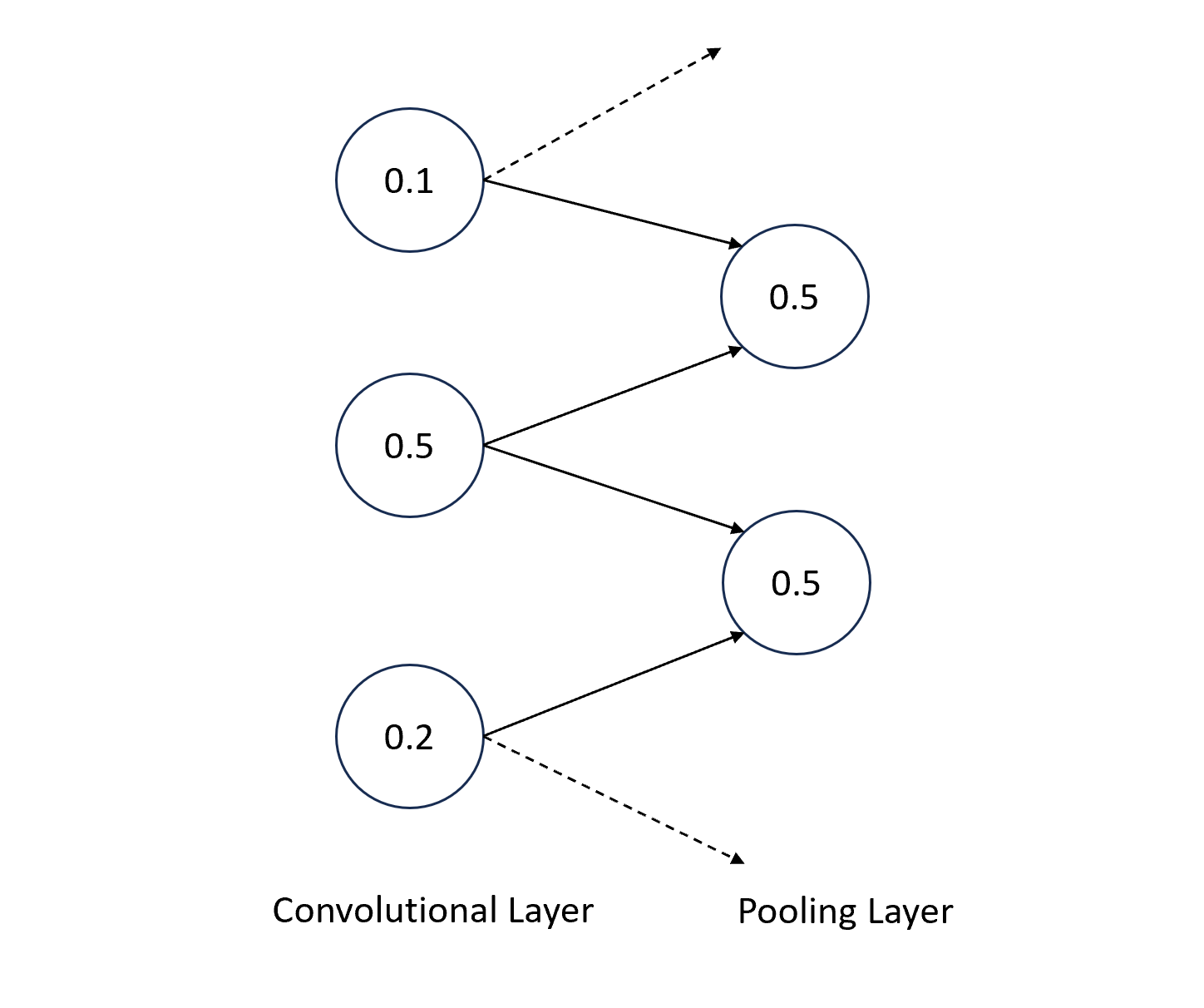}
    \caption{Diagram of Max Pooling Layer. The circles on the left represent the output of a convolutional layer. The circles on the right represent the output from the max-pooling layer, with a stride of 2 neurons and a pooling region width of 2 neurons. Each circle represents a neuron in the layer, and the value is written inside. The neurons in the max pooling layer take the maximum value over the neurons in the convolutional layer that are covered by the pooling region.}
    \label{fig:max pooling}
\end{figure}

We can next consider pooling layers. Pooling layers compress the information output from the previous layer into a lower-dimensional space. We use a specific version of pooling called \textit{max pooling}, where given the output of the previous layer as the input, the output of the pooling layer is a vector of the maximum values of small regions along the input vector. The stride length is how many units apart each region is, and the pooling width is the number of neurons summarized in each region, as can be seen in Figure \ref{fig:max pooling}. Pooling helps to keep the output invariant to small translations of the input, which is useful when we solely want to know if a characteristic is present, rather than its exact location.

\subsubsection{Fully Connected Layers}
Finally, we consider the fully connected layers, where the last layer outputs the final prediction. Every unit in the fully connected layer takes the entire output from the previous layer as the input. The activation is defined by,
\begin{equation} \label{eq: Fully connected layer}
    \mathbf{c}_i = \phi (\mathbf{W}_i \mathbf{c}_{i-1} + \mathbf{d}_i ),
\end{equation}
where $i$ is again the layer number in the CNN, $\mathbf{c}_i$ is a vector of length $n_i$ of activations in layer $i$, $\mathbf{W}_i$ is an $n_i \times n_{i-1}$ matrix of learned weights, $\mathbf{d}_i$ is a vector of length $n_i$ of learned bias parameters, and $\phi$ is an activation function as described in Section \ref{sec:activation functions}.

\subsubsection{Activation Functions} \label{sec:activation functions}
An activation function is a non-linear mathematical operation that determines the output from a unit, introducing non-linearity to enable the network to learn complex patterns and make better predictions. The choice of $\phi$ depends on the application. The most popular activation function, rectified linear unit (ReLu), is given by,
\begin{equation}
    \phi(x) = \mathrm{max} \{0,x\}.
\end{equation}
For many activation functions, the derivative vanishes as $x\rightarrow \pm \infty$, which can lead to slow convergence for gradient-based optimization algorithms. The derivative of the ReLu function on the other hand does not face this issue, and so tends to converge quicker and with better performance. We used ReLu activation functions in our CNN layers.

\subsection{Training the networks}
Neural networks are trained to minimize a loss function. We provide the neural network with a training set containing examples of inputs and true values, and the loss function is a measure of the difference between the prediction from the model, and the true value. The mean squared error (MSE) is a common loss function used for regression tasks, and is given by
\begin{equation}
    L(\hat{y_j}, y_j| \mathbf{p}) = \frac{1}{M} \sum_{j=1}^M (\hat{y_j} - y_j)^2,
\end{equation}
where $y_1,y_2,\dots ,y_M$ are the true labels of all $M$ examples in the training set, and $\hat{y_1},\hat{y_2},\dots ,\hat{y_M}$ are the predicted outputs given parameters $\mathbf{p}$, where $\mathbf{p}$ is the vector of the parameters in the model. The values of $p$ are learned during training. For the CNN, in the convolutional layers these are the elements of all convolutional kernels $\mathbf{w}_n^{(t,l)}$ and bias vectors $\mathbf{b}_n$ from Eq.\ref{eq: CNN layer}, and for the fully connected layers these are the weight matrices $\mathbf{W}_n$ and bias vectors $b_n$ from Eq.\ref{eq: Fully connected layer} where $n$ corresponds to the index of the final output layer. 

We minimize the loss function through an optimization algorithm called gradient descent. To reach this minimum, we first start with a random set of parameters $\mathbf{p}$, and then iteratively update them, so that we descend along the gradient with respect to the parameters. If we write the loss function as $f(\mathbf{x})$ then this can be expressed as,
\begin{equation}
    \mathbf{x}' = \mathbf{x} - \epsilon \nabla_\mathbf{x} f(\mathbf{x}),
\end{equation}
where $\nabla_\mathbf{x} f(\mathbf{x})$ is the gradient, and $\epsilon$ is a positive scale factor called the learning rate, which determines the step size between each iteration. The learning rate is a tunable hyperparameter as further described in Section \ref{subsec: preparing training, validation, test sets}. We tune this rate until we reach a suitable minimum value of the loss function.

An important class of minimization algorithms is called Stochastic Gradient Descent (SGD). To find the true gradient the entire training set of size $M$ must be used. However, this is computationally expensive, and so in SGD a random subset of the data is used of size $B_{\text{SGD}}$, where $B_{\text{SGD}}$ is called the 'batch size', and $1\leq B_{\text{SGD}} << M$. Then the minimization is performed using this approximate gradient. We used a constant batch size of $B_{\text{SGD}}=1024$, as $B_{\text{SGD}}$ does not affect the performance of the model if the other hyperparameters are well tuned, as demonstrated by \cite{shallue2019measuring}. We used a variant of the SGD algorithm, SGD with momentum \citep{POLYAK19641}, and fixed the momentum parameter at $0.9$.

\subsection{Preparing Training, Validation, Test Sets} \label{subsec: preparing training, validation, test sets}
In ML, it is the gold standard to split your data into three subsets: a training, validation, and testing set. This is done to ensure generalization, which is the ability of a model to perform well on new, previously unseen data \citep{Goodfellow-et-al-2016}. A ML model is first trained using the training set, which commonly contains 80\% of the data. The validation set contains data that the model has not seen yet and is used to evaluate the performance of the model on new inputs at every step of the learning process. For example, this can be used to optimize the model architecture. This process allows us to prevent under- and overfitting on the observations, which should be avoided since it can result in poor out-of-sample performance. Finally, after the model has been tuned using the training and validation sets, the final model's performance can be evaluated using the testing set.

Since we are using a relatively small dataset (hundreds of examples compared to the thousands commonly used in ML), we use a $k-$fold cross-validation method on our training data set, as this allows us to make out-of-sample performance for all of the dataset, rather than only on the relatively small validation and test sets. In our model, we split the data as 80\% in training, 10\% in validation, and 10\% in testing. The steps for the $k$-fold cross-validation method performed on the training set are as follows.
\begin{enumerate}
\item The training data is randomly assigned into $k$ subsets.
\item For each training data subset:
    \begin{enumerate}
        \item take the current subset as the `hold out' data set,
        \item take the other $k-1$ groups as the new training data set,
        \item fit a model using the new training data set, and evaluate it using the hold-out data set,
        \item record the model performance.
    \end{enumerate}  
\end{enumerate}
Using this strategy, each of the $k$ subsets is used in the hold-out group once. The value of $k$ must be chosen so that there is not a high variance (lower variance implies model performance changes a lot based on training data, which can occur for higher $k$ values) or a high bias (over-estimation of the model performance occurs for high $k$). We used a 10-fold cross-validation method, as $k=10$ has been shown empirically to result in test error estimates that do not give high bias or variance \citep{James2023StatisticalLearning}. We then optimized our model by first training using the 10-fold cross-validation, and using the 10\% validation set to tune the hyperparameters as detailed in Section 4.6. Finally, we evaluated the optimized model performance on the 10\% test set. 

\subsection{Overfitting and Regularization}
Compared to other optimization algorithms, neural network methods are especially capable of picking up on subtle patterns in datasets. However, they can also be especially prone to overfitting. To prevent this and ensure proper generalization, which is the ability of a model to perform well on new, previously unseen inputs, we implemented regularization methods. The generalization of a model is often described by the training error and test error. The error of predictions on the training set is called the training error, and we try to make this low for the best results to reduce underfitting. We also compute the generalization error or the test error, which is the expected value for the error on a new input, and minimization reduces overfitting. Methods involving changing the learning algorithm to reduce the generalization error but not the training error are called regularization methods. The method we used is weight decay regularization, which reduces the complexity of the model by limiting the values the parameters can take. If we write the parameters as a vector $p$, then on each iteration it updates as,
\begin{equation}
    \mathbf{p}_{i+1} = (1-\varepsilon) \mathbf{p}_i + (\nabla \mathbf{p}_i )_{\text{opt}},
\end{equation}
where $(\nabla \mathbf{p}_i )_{\text{opt}}$ is the change computed by the original algorithm at iteration $i$, and $1-\varepsilon$ is a factor to reduce the parameter vector by. The value for $\varepsilon$ was optimized during hyperparameter tuning.

\subsection{CNN Implementation and Hyperparameter Optimization}
We implemented the CNN model in \texttt{TensorFlow}, an open-source software library for ML \citep{tensorflow2015-whitepaper}. 

To minimize the loss function, we used SGD with momentum on the cross-validation set. To find the optimum hyperparameters for the model, we performed $\sim 300$ random searches across the parameter space over the cross-validation set for the learning rate, kernel size, filters, convolutional, fully connected, and max pooling layers, units, pool size, pool strides, weight decay, and the number of epochs as listed in Table \ref{table:hyper parameter space}. We did this for $\sim10$ different models, with each model containing the original white light CCF data, and one additional feature.

We evaluate a model's performance using the root mean-squared error (RMSE) between the labels and predictions on the validation set. For a more detailed description of the RMSE see Appendix \ref{sec: Appendix RMSE}.
To find the optimum hyperparameters, we first found the hyperparameter values that corresponded to the lowest RMSEs for each of the models, and then found the best value across all the random searches and over all the models, by visualising how the RMSE varied using the \texttt{TensorBoard}\footnote{\url{https://www.tensorflow.org/tensorboard}} software included in the \texttt{TensorFlow} library \citep{tensorflow2015-whitepaper}. These values are listed as the `Optimized Values' in Table \ref{table:hyper parameter space}. These optimized values were then used for our best CNN model architecture.

We then trained several different models using the best set of hyperparameters. The input to the first model trained was only the white light CCF. Then, we included one additional feature in each model, which was added after the max pooling layer as illustrated in Figure \ref{fig:cnn structure flow best model}. The additional features were the white light CCF bisector and FWHM, S-Index, the EW of H$\alpha$ lines and Sodium D1 and D2 lines, the TSI and TSI derivative, the unsigned magnetic flux, and the chromatic CCFs. Then, for the best performing features, we trained models including several of these features simultaneously. In the case of the blue, yellow, and red CCFs, adding one of these would result in 49 features since the CCFs are 49-dimensional arrays rather than 1-dimensional arrays such as features like s-index, EW of H$\alpha$, etc.

\begin{deluxetable}{cccc}

\tablecaption{\label{table:hyper parameter space} Hyperparameters used to train the model, including whether they were discrete or logarithmic (column 2), the values of the tested hyperparameters (column 3), and the optimized values corresponding to the final model used (column 4). To find the optimized hyperparameters for the final model we performed random searches across the parameter space. For the hyperparameters categorized as logarithmic, the learning rate was sampled by generating a range of exponents uniformly and then applying exponentiation to create a log-scale distribution. The weight decay values were generated to be evenly spaced on a logarithmic scale.}

\tablehead{
\colhead{Hyperparameters} & \colhead{Hyperparameter} & \colhead{Random Search} & \colhead{Optimized} \\
[-1.4ex]
& \colhead{Distribution} & \colhead{Space} & \colhead{Values}
}
\startdata
    Learning Rate & Logarithmic & $10^{-4} - 10^{-2}$ & 0.003 \\
    Conv Kernel Size & Discrete & 1,3,5,7,9,11 & 9\\
     No. conv filters & Discrete & 2,4,8,16,32,64 & 8\\
     No. conv layers & Discrete & 1,2,4,6,8,10& 1\\
     No. dense units & Discrete & 50,100, 200 & 500\\
     & & 500, 1000, 2000 \\
     No. dense layers & Discrete & 1,2,4,6,8,10,12 & 4\\
     No. Max Pooling & Discrete & 1, 2, 3, 4 & 1\\
     layers & & \\
     Pool Size & Discrete & 1, 2, 3, 4, 7, 10 & 3\\
     Pool Strides & Discrete & 1, 2, 3, 4, 7, 10 & 2\\
     Weight decay & Logarithmic & $0.0005 - 0.05$ & 0.003\\
     Epochs & Discrete & 50, 55, 65, \\
     & & 70, 80, 90, & 90\\
     & & 100,110,120\\
\enddata
\end{deluxetable}

\section{Results} \label{sec: Results}
\subsection{Performance metrics}
\begin{figure}
    \centering
    \includegraphics[width=\columnwidth]{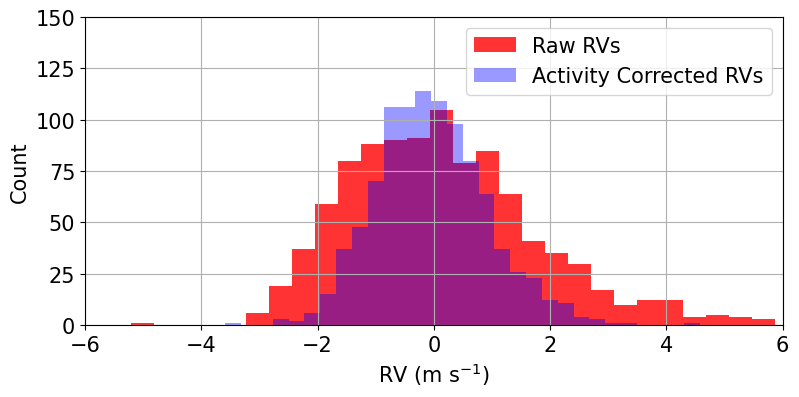}
    \caption{Distribution of RV values before correcting for activity (raw RVs, red)  compared to the activity corrected RVs distribution (blue) using the CNN model. The RV labels are skewed from a standard normal distribution. The stellar activity corrected RVs have a lower variance than the raw RVs, indicating that our neural network is effective in mitigating activity signals.}
    \label{fig:labels preds histograms}
\end{figure}

Although RMSE is an easily accessible metric in our \texttt{Tensorflow} implementation of the CNN and can be a helpful metric to assess model performance,  RMSE is not resistant to outliers in the dataset, and our dataset does have some non-Gaussian characteristics and outliers \footnote{The non-Gaussian RV distribution arises primarily from stellar activity, as magnetic active regions suppress convective blueshift, leading to positive RV deviations, and thus causes a skew in the data.} (see Figure \ref{fig:labels preds histograms}). Therefore to determine which stellar activity features work the best in the CNN in providing predictions of the RV, we calculate the metric $\sigma_\mathrm{percentile}$ instead. We do this using the 84th and 16th percentiles, which we describe in detail in Appendix \ref{sec: Appendix sigma percentile}.

We calculate $\sigma_\mathrm{percentile}$ for both the raw RVs (without any activity mitigation), which correspond to the RV labels used in the neural network, and the ``corrected" RVs, $\mathrm{RV}_\mathrm{corrected}$, where we have removed the predicted activity signal from our neural networks. In particular, we define: 
\begin{equation} \label{eq: corrected RVs}
    \mathrm{RV}_\mathrm{corrected} = \mathrm{RV}_\mathrm{labels} - \mathrm{RV}_\mathrm{predicted}
\end{equation}

To determine which combination of features produces the best stellar variability model, we trained $100$ different instances of each model architecture, where each instance is initialized with different random values for each parameter before optimization. We chose to train such a large number of models because any given architecture can perform better or worse depending on the input initialization, which will vary due to randomness in data ordering during training. By training many models and averaging the results, we can ensure that our model architecture comparisons are robust to these random variations.

For each of the $100$ models trained for each architecture, we find the $\sigma_\mathrm{percentile}$ for the corrected RVs. Then, we calculate the median and standard error of the median (see Appendix \ref{sec: Appendix standard error on median} for details) over the $100$ models. By using these metrics for each model architecture we can robustly and fairly compare each model's performance.

\subsection{ML Results}
\begin{figure}
    \centering
    \includegraphics[width=\columnwidth]{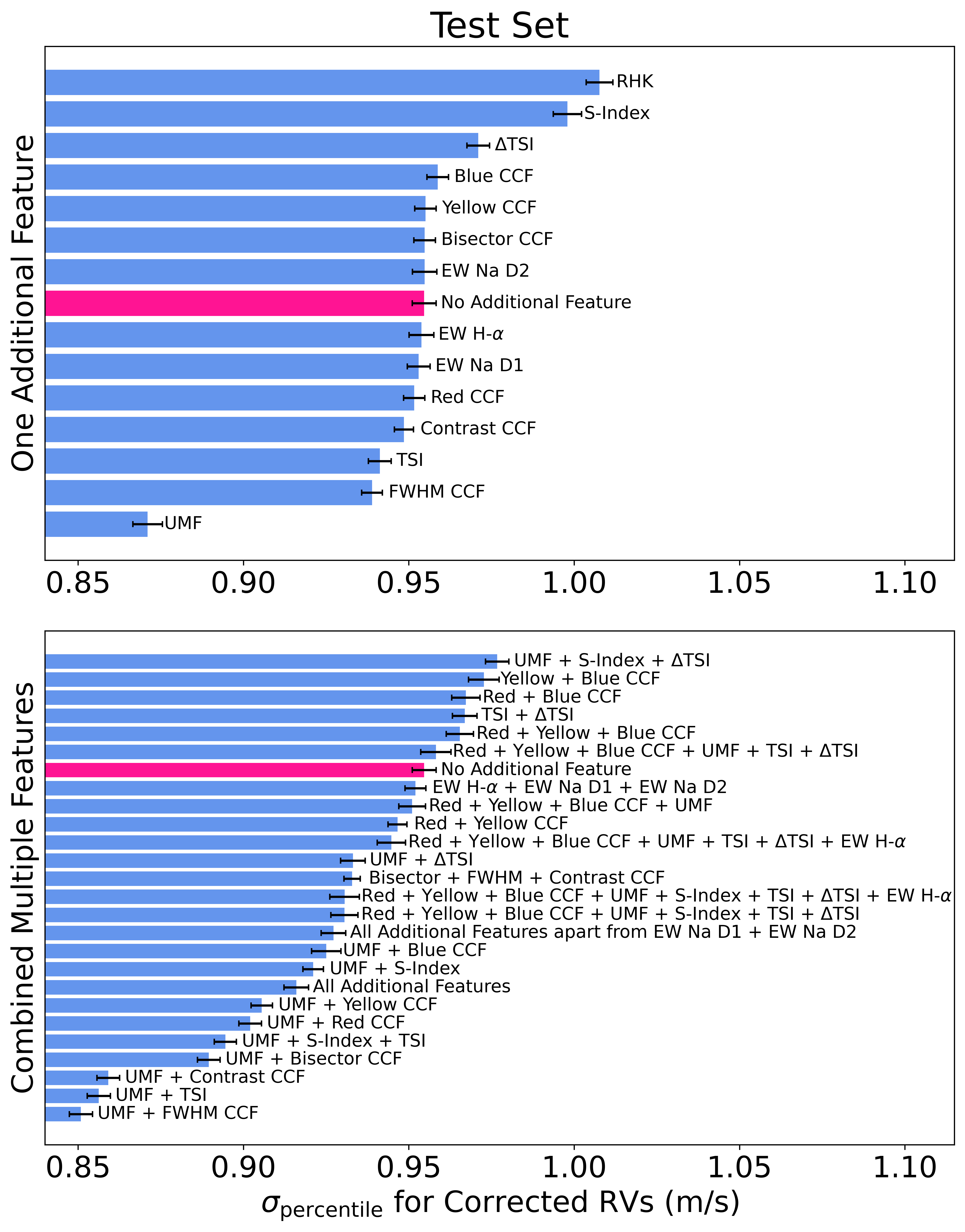}
    \caption{Median $\sigma_{\text{percentile}}$ over $100$ runs for the corrected RVs over the test set, for different model architectures. The standard errors of each median are shown as black error bars. Models have either one additional feature to the original white light CCFs (top) or multiple additional features combined (bottom).
    The model in pink uses only the original white light CCFs as the input to the CNN, and the models in blue represent those with any additional features.
    There are abbreviations for unsigned magnetic flux (UMF), Total Solar Irradiance (TSI), and TSI Derivative ($\Delta$TSI). The most effective parameters for the NN are the UMF, TSI, and $\Delta$TSI.}
    \label{fig:bar chart results full sigma}
\end{figure}

The median and uncertainty on the $\sigma_\mathrm{percentile}$ scatter in the corrected RV time series for each model architecture are shown in Figure \ref{fig:bar chart results full sigma} and Table \ref{table:sigma percentile values test set}. The Figure shows the results across the test set. It includes models using only one feature in addition to the white light CCFs, as well as models with multiple additional features. Table \ref{table:sigma percentile values test set} focuses on models with multiple additional features. Additional Figures and Tables comparing results across the cross-validation, validation, and full datasets individually for both RMSE and $\sigma_\mathrm{percentile}$, for all model architectures, can be found in the Appendices (Figure \ref{fig:bar chart results}, Table \ref{table:rmse values}, Figure \ref{fig:bar chart results sigma percentile}, Table \ref{table:sigma percentile values}).

Overall, while the relative performance of the models vary slighlty depending on which datasets and scatter metrics are used for evaluation, a few consistent patterns emerge. In particular, we see that the unsigned magnetic flux, TSI, TSI derivative, chromatic CCFs, S-Index, H$\alpha$ EW, and the contrast and FWHM of the white light CCFs consistently reduce the $\sigma_\mathrm{percentile}$ scatter most compared to the models with no additional features added (only the white light CCFs are input). We also note that the bisector of the white light CCFs, and EW of the Sodium D1 and D2 absorption lines do not seem to improve the model performance, suggesting that they may not contain any additional information that is not already captured by the white light CCFs.

\begin{deluxetable}{cc} 

\tablecaption{\label{table:sigma percentile values test set} The median and uncertainty on the $\sigma_{\text{percentile}}$ scatter in the corrected RVs for 100 models run for different architectures over the test set. Only architectures that use more than one additional feature are included. They are in descending order of median $\sigma_{\text{percentile}}$ for the test set, with lower values indicating improved performance. There are abbreviations for unsigned magnetic flux (UMF), Total Solar Irradiance (TSI), and TSI Derivative ($\Delta$TSI).}

\tablehead{
\colhead{Additional Input Features} & \colhead{$\sigma_{\text{percentile}}$} 
\\ [-1.4ex]
\colhead{} & \colhead{(\cms)}} 

\startdata
      UMF + S-Index + $\Delta$TSI & $97.67\pm0.35$\\
      Yellow CCF + Blue CCF &  $97.27\pm0.46$\\
      Red CCF + Blue CCF  & $96.73\pm0.43$\\
      TSI + $\Delta$TSI & $96.69\pm0.37$\\
      Red CCF + Yellow CCF + Blue CCF & $96.54\pm0.41$\\
      Red CCF + Yellow CCF + Blue CCF  & $95.82\pm0.46$\\
      + UMF + TSI + $\Delta$TSI & \\
      No additional features  & $95.46\pm0.36$\\
      H$\alpha$ EW + Sodium D1 EW  & $95.20\pm0.32$\\
      + Sodium D2 EW & \\
      Red CCF + Yellow CCF + Blue CCF  & $95.10\pm0.40$\\
      + UMF & \\
      Red CCF + Yellow CCF &  $94.66\pm0.29$\\
      Red CCF + Yellow CCF + Blue CCF  & $94.47\pm0.43$\\
      + UMF + TSI + $\Delta$TSI + H$\alpha$ EW & \\
      UMF + $\Delta$TSI & \textbf{$93.31\pm0.37$}\\
      CCF Contrast + CCF Bisector + CCF FWHM & $93.29\pm0.25$\\
      Red CCF + Yellow CCF + Blue CCF & $93.06\pm0.45$\\
      + UMF + S-Index + TSI  & \\
        + $\Delta$TSI + H$\alpha$ EW & \\
      Red CCF + Yellow CCF + Blue CCF &  $93.05\pm0.41$\\
       + UMF + S-Index + TSI + $\Delta$TSI & \\
      UMF + S-Index + TSI + $\Delta$TSI + H$\alpha$ EW & $92.72\pm0.37$\\
      + Red CCF + Blue CCF + Yellow CCF \\
      + CCF Contrast + CCF Bisector + CCF FWHM \\
      UMF + Blue CCF & $92.51\pm0.44$\\
      UMF + S-Index & $92.11\pm0.31$\\
      UMF + S-Index + TSI + $\Delta$TSI & $91.60\pm0.37$\\
      + Red CCF + Blue CCF + Yellow CCF \\
      + H$\alpha$ EW + Na D1 EW + Na D2 EW\\
      + CCF Contrast +  CCF Bisector + CCF FWHM\\
      UMF + Yellow CCF &  $90.55\pm0.32$\\
      UMF + Red CCF &  $90.21\pm0.34$\\
      UMF + S-Index + TSI & $89.45\pm0.34$\\
      UMF + CCF Bisector & $88.95\pm0.34$\\
      UMF + CCF Contrast & $85.91\pm0.34$\\
      UMF + TSI &  $85.63\pm0.35$\\
      UMF + CCF FWHM & $85.09\pm0.35$\\
\enddata

 \end{deluxetable}

\begin{figure}
    \centering
    \includegraphics[width=\columnwidth]{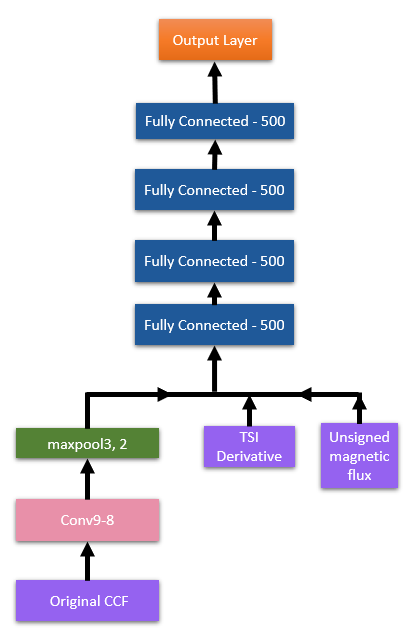}
    \caption{Architecture of our best performing CNN model. This has both optimized hyperparameters and optimized input features. Convolutional layers area denoted Conv$\langle$Kernel Size$\rangle$-$\langle$No. conv filters$\rangle$, max pooling layers are denoted maxpool $\langle$Pool Size$\rangle$-$\langle$Pool Strides$\rangle$, and fully connected layers are denoted Fully Connected $\langle$No. dense units$\rangle$. The features being input into the CNN are shown in purple.}
    \label{fig:cnn structure flow best model}
\end{figure}

\begin{figure*}
    \centering
    \includegraphics[width=\textwidth]{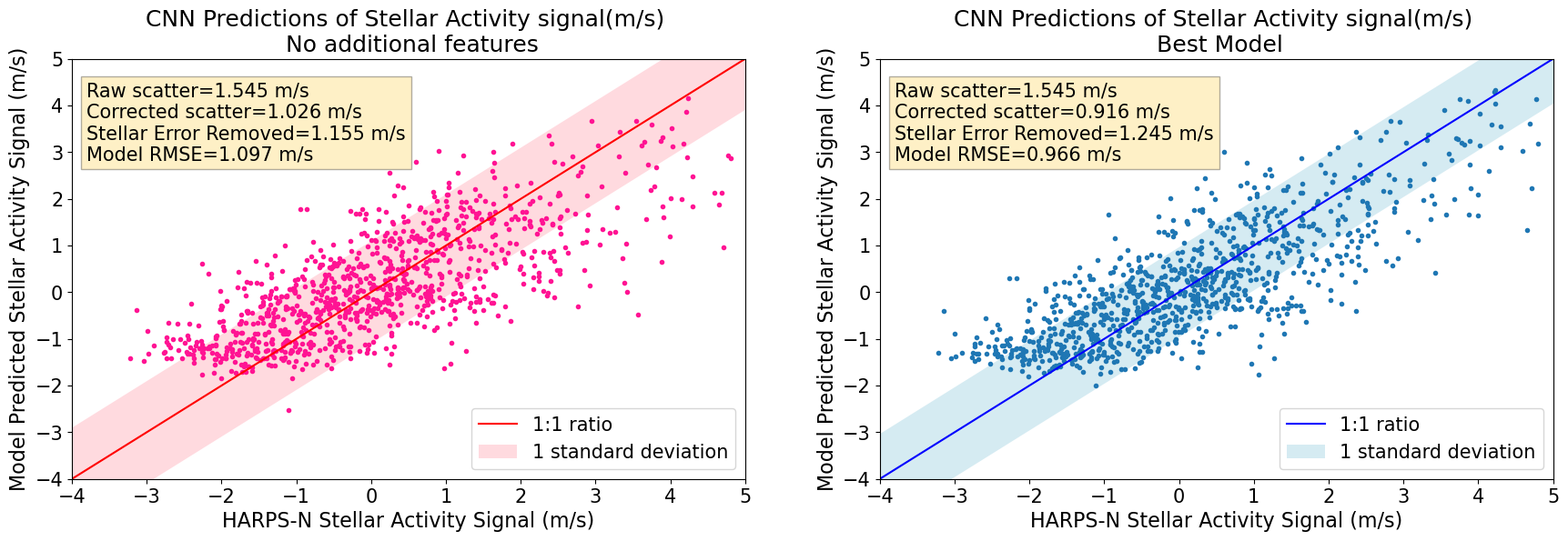}
    \caption{Comparison between the observed RVs and the predictions from the CNN model. The raw and corrected scatter are given by $\sigma_{\text{Percentile}}$ for the $\text{RV}_{\text{labels}}$ and $\text{RV}_{\text{corrected}}$ respectively, and model RMSE is the root mean square error of the $\rm RV_{\rm corrected}$. The stellar error removed is the difference between the raw scatter and corrected scatter in quadrature. Left: The predictions are from a model with the best hyperparameter architecture, but with no additional features. Right: The predictions are from a model with the best hyperparameter architecture, and the best features.}
    \label{fig: labels predictions plot}
\end{figure*}

\subsection{Improvement in RV Scatter for a Nominal ``Best'' Model} \label{rvscatterimprovements}

\begin{figure*}
    \centering
    \includegraphics[width=0.85\textwidth]{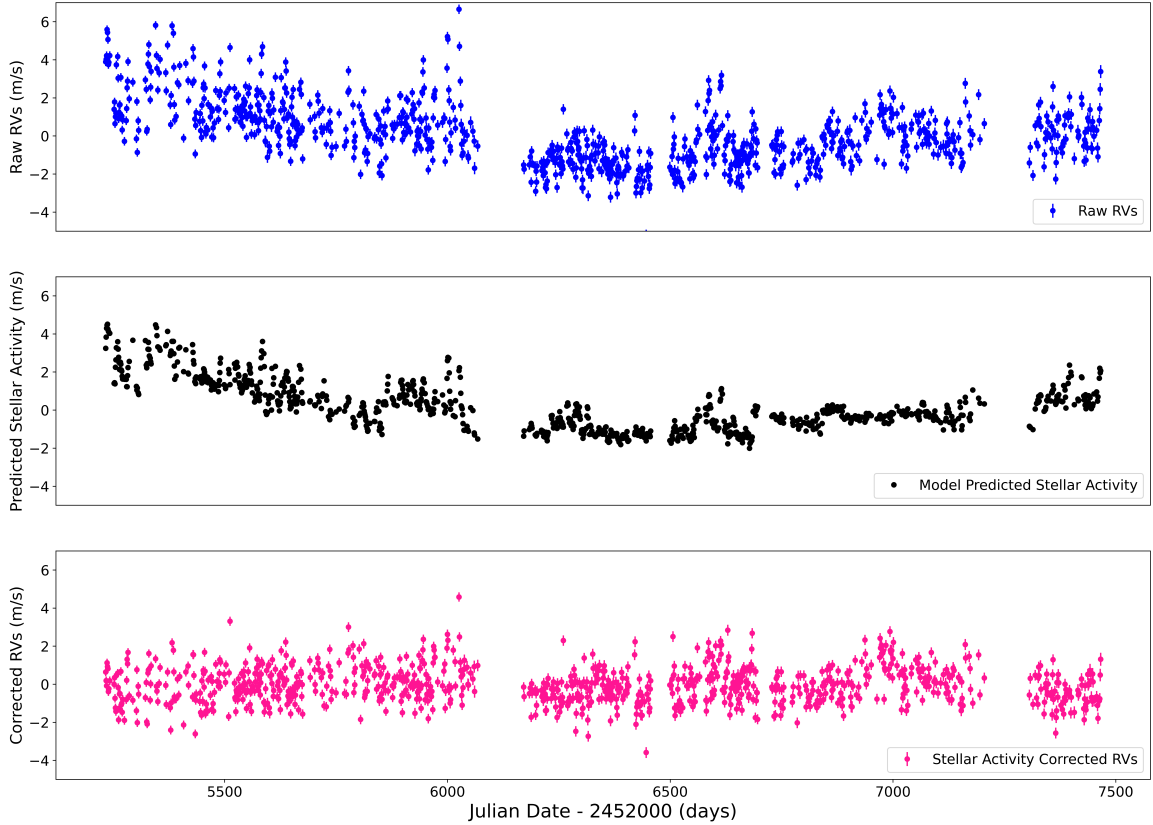}
    \caption{Time series of the HARPS-N radial velocity measurements and our neural network's activity correction. Top: Uncorrected HARPS-N data. Middle: CNN-predicted activity signals from our best architecture. Bottom: HARPS-N radial velocity measurements corrected by subtracting the activity predictions. The gaps in the observations at $\sim7,200$ days correspond to hardware downtime. RV uncertainties in the top and bottom panel are from the HARPS-N pipeline.}
    \label{fig:all 3 rvs}
\end{figure*}

Many of the architectures we considered performed similarly well, so it is not possible to truly identify which particular features are optimal. Nevertheless, for the sake of simplicity in understanding the results of our tests, we identify a nominal ``best-performing'' model and use its predictions in our analysis for the rest of this paper. We chose the model with the lowest $\sigma_{\text{percentile}}$ value for the corrected RVs in the full dataset (shown in Table \ref{fig:bar chart results full sigma}). This best model uses the unsigned magnetic flux and TSI derivative as additional features, and reduces the  $\sigma_{\text{percentile}}$ from 154.5 \cms to 91.6 \cms. The architecture of the final CNN using this model is shown in Figure \ref{fig:cnn structure flow best model}. The results of this best model across the full dataset are summarized in Figure \ref{fig: labels predictions plot}. For our best architecture for a model with no additional features (i.e. only the white light CCFs), we find that the corrected RV scatter is 102.6 \cms. However, adding in the TSI derivative and unsigned magnetic flux improves the results significantly; we measure a corrected RV scatter of 91.6 \cms. Similarly across the test set, for a model with no additional features the corrected RV scatter reduces from 147.1 \cms\ to 95.5 \cms , and using the best model this reduces to 93.3 \cms. This highlights that these additional features further help the CNN to be able to model and predict the RV signals originating from stellar activity better. For the best model architecture, the raw RVs, CNN predicted stellar activity contributions, and CNN stellar-activity corrected RVs are plotted over time in Figure \ref{fig:all 3 rvs}.

\subsection{Improvement in Radial Velocity Scatter in Fourier Space}\label{sec:periodogram}
\begin{figure*}
    \centering
    \includegraphics[width=0.85\textwidth]{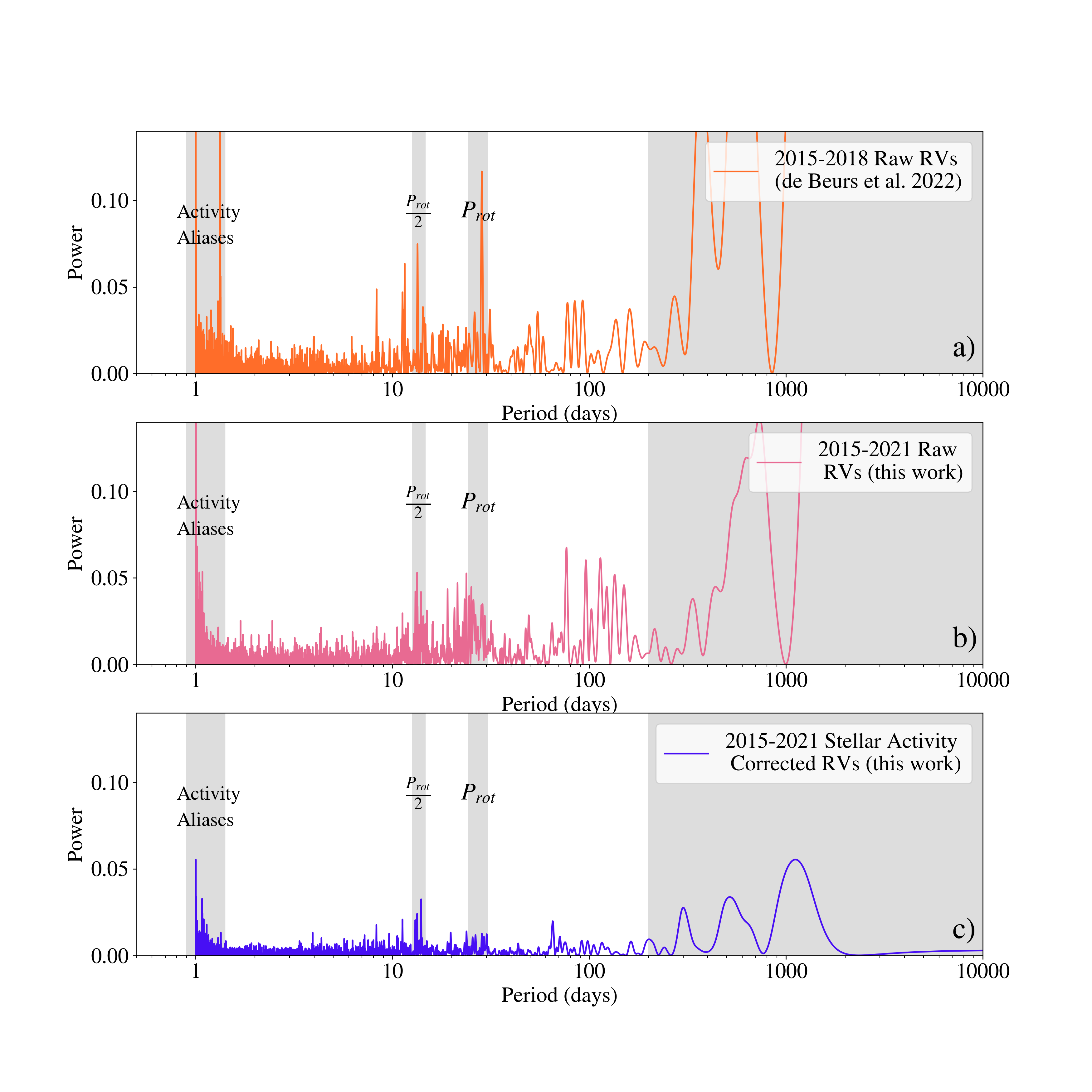}
    \caption{Lomb-Scargle periodograms of the HARPS-N RVs used in this work and in \citet{deBeurs2022}. a) The three years of raw RVs included in \citet{deBeurs2022} in Fourier space. b) The 6 years of raw RVs in Fourier space included in this work. The peak at the rotation period is less strong in this work, which is likely because our dataset includes a larger fraction of observations collected during solar minimum (see Section \ref{sec:periodogram}). c) The full six-year dataset after subtracting our best neural network's activity prediction. The overall activity signal in the periodogram decreases, with particularly noticeable decreases in the amplitude of the periodogram peaks corresponding to the stellar rotation period ($\rm P_{\rm rot}$) and long-term activity cycle.}
    \label{fig:periodogram}
\end{figure*}

To understand which RV signals are being modeled and removed by the best model, and whether these signals correspond to stellar activity, we examined the RVs in the Fourier domain. The raw and corrected RVs are shown in Figure \ref{fig:periodogram} in the form of a Lomb-Scargle periodogram \citep{Lomb1976,Scargle1982}. To obtain the Lomb-Scargle periodogram we used the \texttt{LombScargle} \footnote{\url{https://docs.astropy.org/en/stable/timeseries/lombscargle.html}} function included in the \texttt{astropy.timeseries} package \citep{VanderPlas2015, astropy:2013, astropy:2018, astropy:2022}. We used the periodogram normalization outlined in \citet{Zechmeister2009}.

The peaks indicated in Figure \ref{fig:periodogram} as $P_{\rm rot}$ and $P_{\rm rot}/2$ correspond to the Sun's mean synodic Carrington rotation period of $\sim$ 27 days and half this rotation period respectively. The peaks at periods $>200$ days correspond to long-term magnetic cycles. The left-most peaks correspond to aliases resulting from sampling of the RVs once a day. 

A similar Lomb-Scargle periodogram was produced in \citet{deBeurs2022} using only 3 years of HARPS-N data, which we include here in Figure \ref{fig:periodogram}a. For this subset of the data, the peaks in the raw RVs at $P_{\rm rot}$ and $P_{\rm rot}/2$ appear much stronger compared to our full six years of data in Figure \ref{fig:periodogram}b. This is likely due to us using a dataset which is twice as long. For the 3-year dataset, the spots that we see at the rotation period coming in and out of view at that cadence will be relatively stable. However, for the 6-year dataset, with twice as much longitudinal data, it is more likely that there are spots which are not stable and could be out of phase, weakening the rotation signal. There is also more time at solar minimum within our larger data set, and this is likely the primary cause of the weakened rotational signal strength.

The corrected RVs from the CNN model in the bottom panel no longer have the peaks that correspond to these stellar activity signals. These results demonstrate that the CNN can detect and remove the quasi-periodic variability arising from spots and plague networks using only the white light CCF, the unsigned magnetic flux, and the TSI derivative, and no information about the timing of the observations.

\subsection{Injection-Recovery Tests}
\begin{figure*}
    \centering
    \includegraphics[width=1\textwidth]{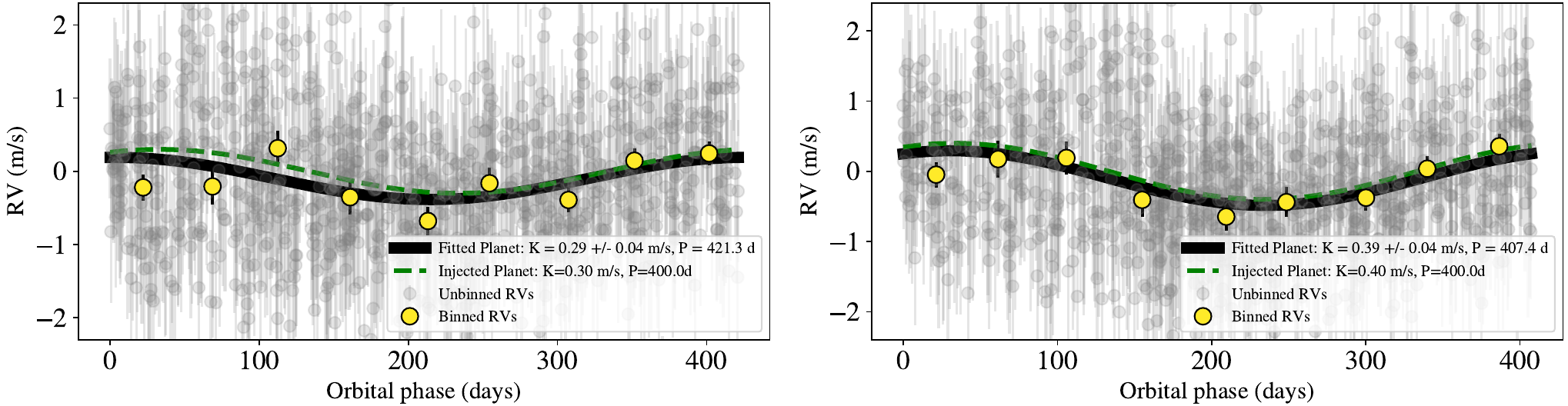}
    \caption{Stellar activity corrected radial velocities with two different injected planetary signals. In the left panel, a planet with semi-amplitude of 0.3\ms on a 400 day orbit is injected into the radial velocities. On the right panel, a planet with a semi-amplitude of 0.4\ms with the same orbital period is used. For both panels, the radial velocities, which are shown with grey points, are phase-folded on the simulated planets' respective orbital periods retrieved by our MCMC analyses. For visualization, the RVs are also binned and labeled in yellow. The injected planet is indicated by a green dotted line and the MCMC retrieved planet is shown by a black line.}
    \label{fig:injection_recovery_phasefolded}
\end{figure*}

Our primary goal in developing our CNN stellar variability model is to improve our ability to detect planetary doppler shifts in radial velocity datasets. To demonstrate that our method preserves planetary signals while modeling stellar variability, we performed several end-to-end planet injection-recovery tests. In these tests, we injected simulated Keplerian planetary signals into the solar data at the CCF level, pass these data through our full pipeline, and then evaluate how well we can recover the orbital parameters of the simulated planets.

Using this approach, we performed four independent planet-injection recovery tests spanning a range of RV semi-amplitudes. For simplicity, we assumed circular orbits with orbital periods of $P=$400 days and with semi-amplitudes of $K=$ 0.10 \ms, 0.20 \ms, 0.30 \ms, and 0.40 \ms. We simulate the planetary RV signals using \texttt{radvel}'s \texttt{kepler.rv\_drive} function \citep{RadVel_fulton_2018} and inject these planetary signals into our CCFs and our HARPS-N radial velocities. Although our best NN model includes unsigned magnetic flux data and TSI derivatives, we do not inject planetary signals into these datasets because we expect unsigned magnetic flux to be unaffected by the presence of planets and only ~0.1-10\% of exoplanets transit (depending on orbital orientations). We do no want to limit our model to only be able to detect transiting exoplanets so we intentionally do not inject planetary transits in our TSI measurements.

To prepare our CCFs and RVs for the neural network model and inject the planet signals into them,  we follow the same steps as described in Section \ref{sec:ccfs}, with the modification that we inject the planetary signal between Step 2 and Step 3 and perform one extra step after Step 3. For the CCFs, the planet signal is injected by applying a translational Doppler shift to the CCFs and for the RVs, the planetary signal is simply added to the HARPS-N RV values. As part of our normal pipeline, we then perform Step 3 described in Section \ref{sec:ccfs}, which is to shift all CCFs to a common center by fitting each CCF to a Gaussian and extracting the $\mu$ values from the Gaussian fit. Notably, this measures and removes translational shifts and is how we would approach any dataset with unknown planetary signal(s). After Step 3, we perform one extra step for the injection-recovery process, which is to subtract the $\mu$ value from our RV values so that the RV values used to train our neural network do not contain translational shift information and focus on stellar variability contributions. Next, we proceed with Step 4 of our pipeline and finally train our neural network on the CCFs, TSI values, unsigned magnetic flux values, and the RVs. After training, we apply the neural network model's stellar activity correction to the RVs to get our corrected RV timeseries. We add our $\mu$ values back to the now stellar activity corrected RVs since those are our estimates of planetary reflex motion.

 Next, we compute Lomb-Scargle periodograms on the corrected RVs with the planetary doppler motion. We require that for a planet signal to be detected, its peak in the periodogram must exceed the False Alarm Probability (FAP) of 0.1\%. With this criteria, we found that the signals corresponding to semi-amplitudes of K=0.30\ms\ and 0.40\ms\ on 400 day orbits produce significant peaks in the periodogram and the lower amplitude signals are not detected. We performed a Monte Carlo Markov Chain (MCMC) orbital fit for both planets using the software \texttt{edmcmc} \citep{vanderburg2021_edmcmc} which incorporates differential evolution to speed up the MCMC fitting process. We applied uniform priors to the period, semi-amplitude, and phase that were broadly centered around the injected values. We fixed the eccentricity to zero. A comparison of the injected planetary orbit and MCMC fit are shown in the left and right panel of Figure \ref{fig:injection_recovery_phasefolded} for K=0.30\ms\ and K=0.40\ms\ respectively. For both planets, the semi-amplitudes estimated from the MCMC, $\textrm{K}_{0.3}={0.29}_{-0.04}^{0.04}$\ms\ and $\textrm{K}_{0.4}={0.39}_{-0.04}^{0.04}$\ms\ agree within error with the true values. However, as seen in the left panel of Figure \ref{fig:injection_recovery_phasefolded}, there is a mismatch between the injected and retrieved planet curves for the 0.30\ms planet, which is due to the MCMC best-fit value of the period ($\textrm{period}_{0.3}={421.3}_{-13.7}^{+30.58}$ days) being overestimated. We therefore consider the K=0.30\ms case a weak detection. The lowest amplitude signal we are able to confidently retrieve is the K=0.4\ms\ at a 400 day orbital period, which corresponds to a 5.16\me\ planet around the Sun.

\section{Discussion} \label{sec: Discussion}
\subsection{Visualizing our features during low and high-magnetic activity periods}
\begin{figure*}
    \centering
    \includegraphics[width=0.85\textwidth]{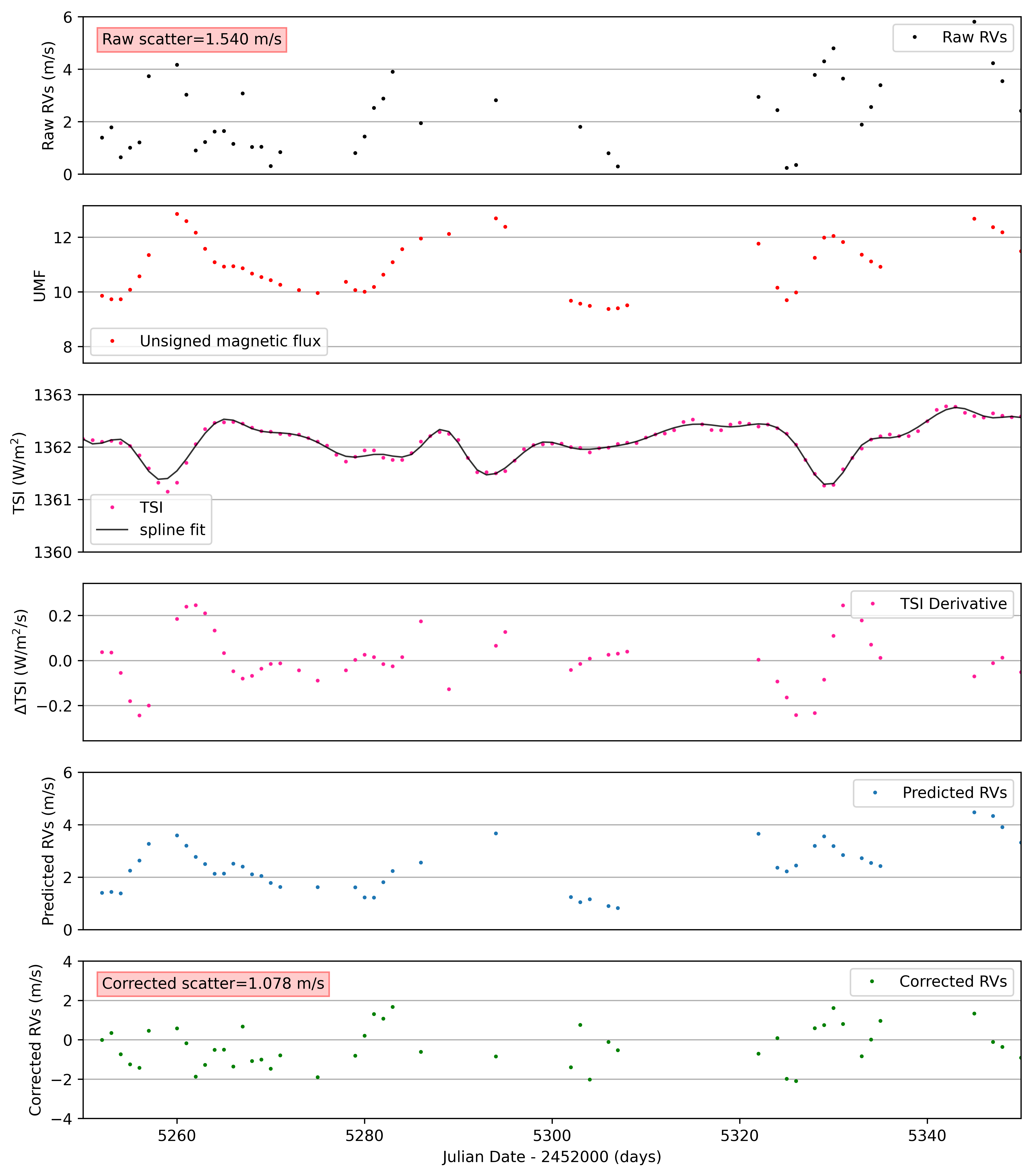}
    \caption{A period of high magnetic activity is shown, with the date 5300 (2457300 JD) corresponding to 2015-10-04, where in this month there was a mean of around 64 sunspots. The RVs from HARPS-N (top panel), unsigned magnetic flux (second), and TSI and TSI spline fit (third), predicted RVs from our best CNN model (fourth), and corrected RVs as given in Equation \ref{eq: corrected RVs} (last panel) are plotted against the Julian date. We see variations at the same times in all of the time series. While most of the features have variations that follow a similar shape to the raw RVs, the TSI is inverted. The value of $\sigma_{\text{percentile}}$, as given in Equation \ref{eq: sigma percentile}, for the raw RVs and the corrected RVs are also shown.}
    \label{fig:high activity}
\end{figure*}

\begin{figure*}
    \centering
    \includegraphics[width=0.85\textwidth]{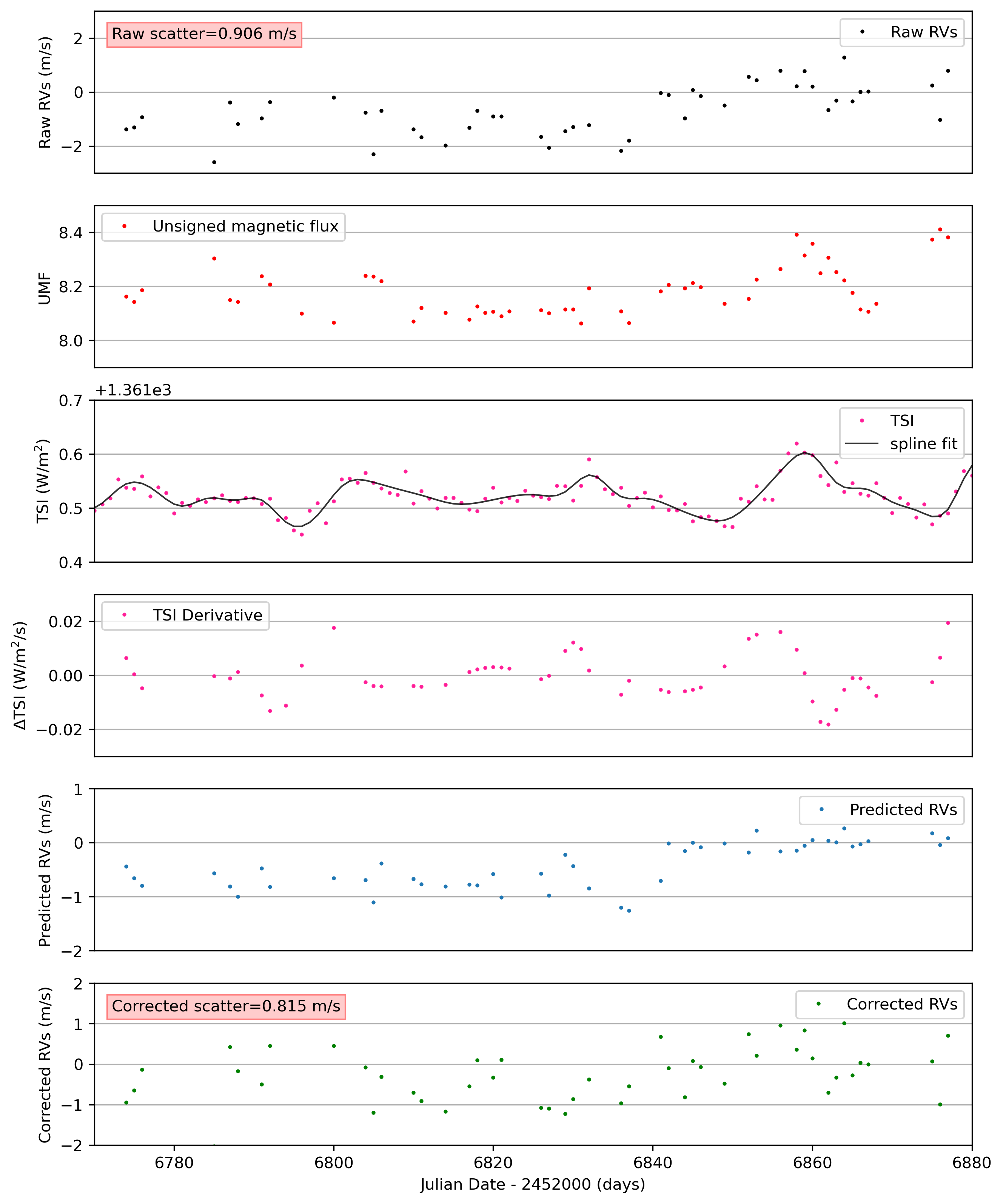}
    \caption{A period of low magnetic activity is shown, with the date 6820 (2458820 JD) corresponding to 2019-12-02, where in this month there was a mean of around 1.5 sunspots. The RVs from HARPS-N (top panel), unsigned magnetic flux (second), and TSI and TSI spline fit (third), and predicted RVs from our best CNN model (fourth), and corrected RVs as given in Equation \ref{eq: corrected RVs} (last panel) are plotted against the Julian date. The value of $\sigma_{\text{percentile}}$, as given in Equation \ref{eq: sigma percentile}, for the raw RVs and the corrected RVs are also shown.}
    \label{fig:low activity}
\end{figure*}

\begin{figure*}
    \centering
    \includegraphics[width=1\textwidth]{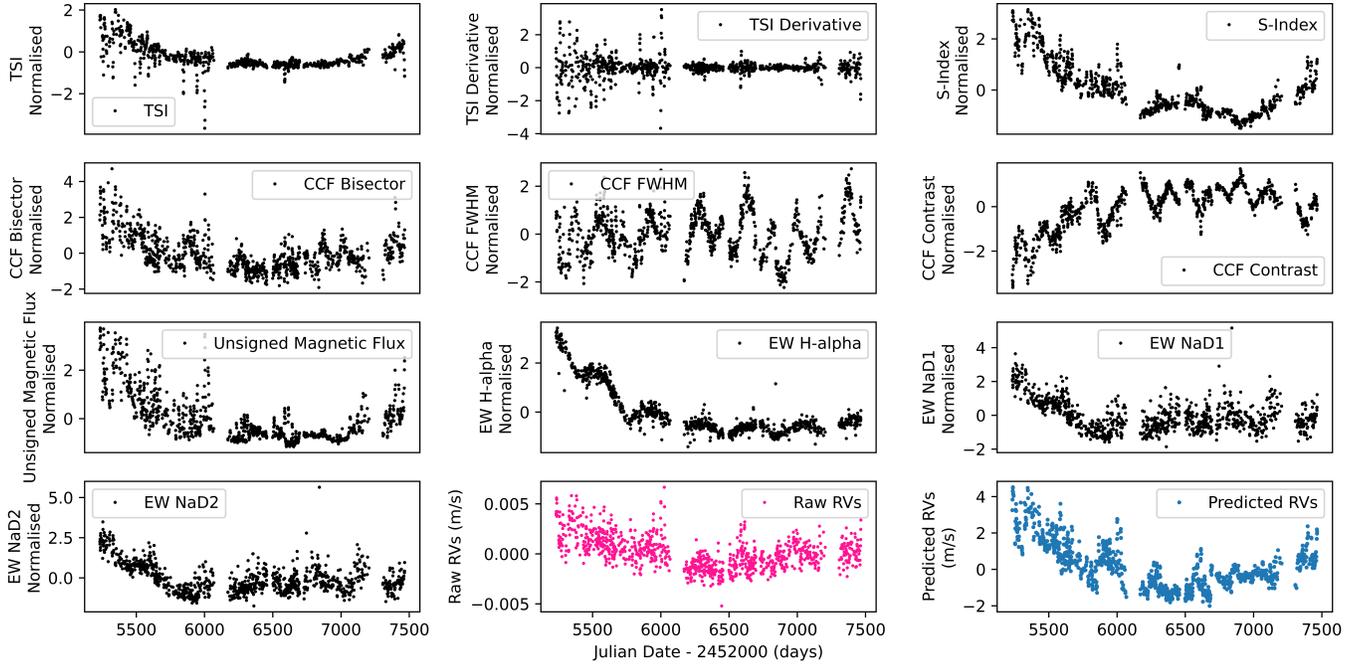}
    \caption{Time series of all of the one-dimensional additional input features to our CNN (shown in black), along with the input raw HARPS-N RVs (pink), and our best model's activity predictions (blue).}
    \label{fig:all features}
\end{figure*}

To qualitatively understand how the magnetic flux and TSI derivative inform the neural network's predictions of stellar variability, we plot in Figures \ref{fig:high activity} and \ref{fig:low activity} the time series for the values of the unsigned magnetic flux, TSI, TSI derivative, raw RVs, predicted RVs, and corrected RVs from our best model during periods of low and high magnetic activity on the Sun. We focus on a period of 100 days, centered around months with one of the highest and lowest number of sunspots in our dataset respectively. The high activity period occurred on October 2015, with a mean of $\sim 64$ sunspots over the month. The low activity period occurred on December 2019, with a mean of $\sim 1.5$ sunspots \citep{SILSO}.

For the high-magnetic activity plot shown in Figure \ref{fig:high activity}, we can see that the unsigned magnetic flux follows the variations in the raw RVs, with similar peaks at around the dates 5260, 5295 and 5330, which correspond to 2457620 Julian Date (JD), 2457295 JD, and 2457330 JD respectively. This is expected due to an increase in magnetic flux during active periods. The TSI instead dips around these dates, indicating a decrease in brightness due to sunspots. The TSI derivative similarly has shape changes at the same positions as the raw RVs, with the most prominent at around the dates 5260 and 5330 on the Figure, which correspond to 2457260 JD and 2457330 JD respectively. We can see that the predicted RVs from our best model capture these RV variations well, with predicted peaks at the same dates. However, the amplitude of the peaks predicted by our neural network are often lower compared to the raw RVs, which is likely due to our dataset mostly containing periods of low solar activity. We suspect that the neural network struggles with predicting the amplitude of the largest activity signals because there are not many examples of such an active sun in the training set. We anticipate that the network's ability to predict these high-activity time periods will improve if we increase the number of high activity examples in the training set. The value of $\sigma_{\text{percentile}}$ for the raw RVs in this period is 154.0 \cms\, and for corrected RVs is 107.8 \cms\, leading to a reduction of 30.0\%.

For the low-magnetic activity time period shown in Figure \ref{fig:low activity}, the predicted RVs follow the amplitude of the raw RVs well, where the value of $\sigma_{\text{percentile}}$ for the raw RVs is 90.6 \cms\ and for the corrected RVs is 81.5 \cms\, leading to a reduction of scatter of 10.0\%. The network is performing well even when there are no strong activity signals in the dataset, and knows to ignore the information from the features (the unsigned magnetic flux, TSI, etc), leading to predicting very low activity signals. Interestingly, the network predicts a jump in the RV time series at Julian date 2458840. This is likely a correction for an instrumental systematic (see Section \ref{instrumentalcorrections}).

In addition, we visualize the entire data set for each of the one dimensional features in Figure \ref{fig:all features}. The raw RVs are shown in pink. We can see that some of the features that did not perform well, such as white light CCF bisector span and white light CCF contrast do not appear to correlate well with the variations in the raw RVs. Bisector span is known to be particularly sensitive to spots. However, over the course of the last three years included in this analysis, the sun was mostly in solar minimum and had few spots and plagues, so this may potentially explain the low correlation of the bisector span with the RVs. On the other hand, features that perform well such as unsigned magnetic flux, TSI, TSI derivative, and s-index have similar overall shape and patterns to the raw RVs. The predicted RVs from our best model are shown in blue, and capture the variations of the raw RVs well, especially at periods of high stellar variability.

\subsection{Instrumental Systematics}\label{instrumentalcorrections}
\begin{figure*}
    \centering
    \includegraphics[width=0.85\textwidth]{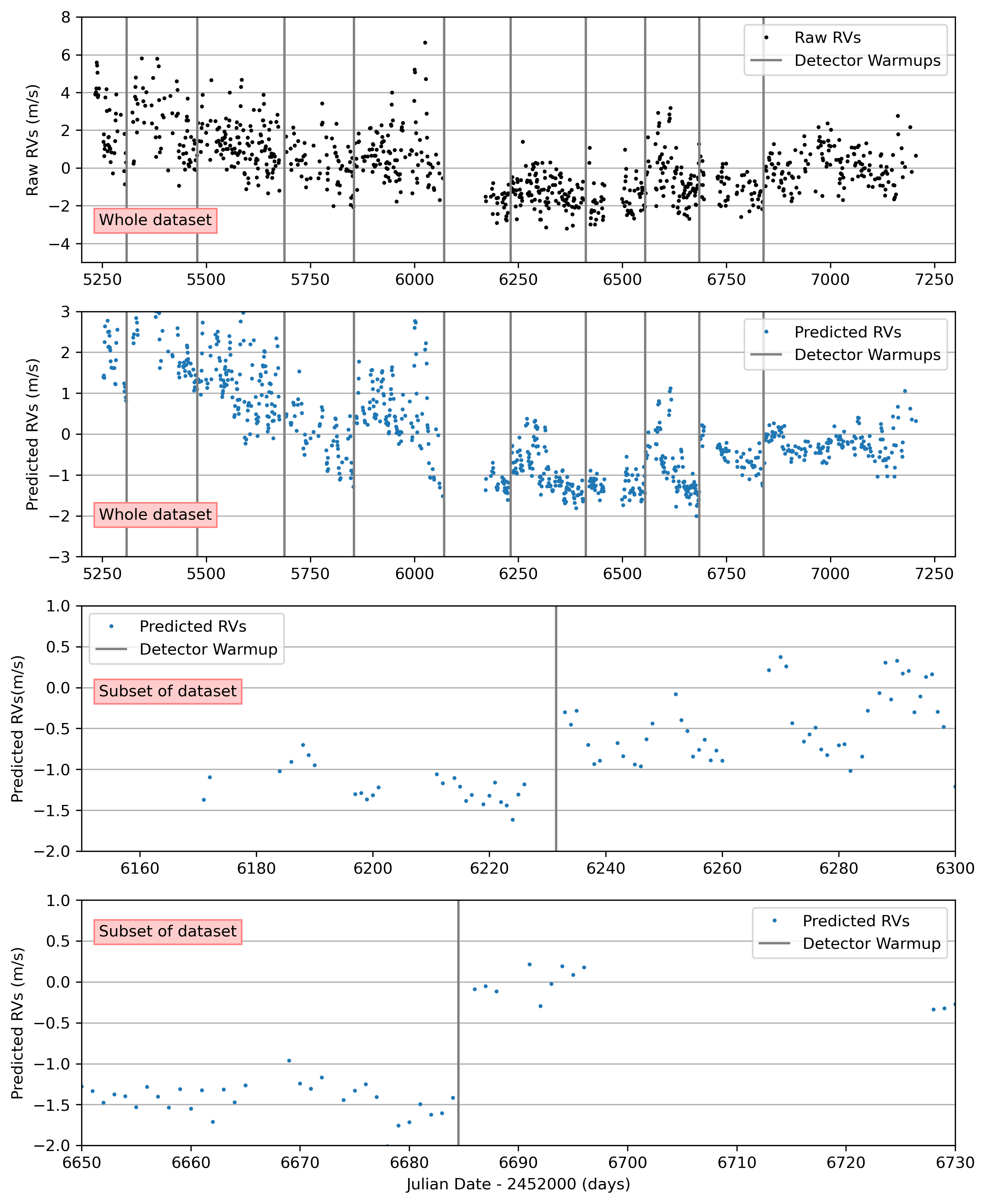}
    \caption{Visualization of the neural network's ability to identify and correct instrumental artifacts. The HARPS-N spectrograph underwent cryostat warm-up events between 2015 and 2021, resulting in temperature jumps in the sensors. The raw RVs (top panel) and predicted RVs from our best CNN model (second panel) are shown, with the warm-up dates overlaid. During periods of higher magnetic activity (first half of the dataset), the effects of warm-ups on the predicted RVs are obscured, but during periods of lower magnetic activity (second half of the dataset), the dates of the warm-ups align with jumps in the predicted RVs. The bottom two panels show subsets of the observations where warm-ups during low magnetic activity periods coincide with increases in predicted RVs. These occur at Julian Date 2458232 (third panel) and Julian Date 2458684 (fourth panel).}
    \label{fig:detector warmups}
\end{figure*}

Although our main goal in developing this neural network was to correct astrophysical noise sources, we found that it was also able to correct instrumental artifacts. In particular, the network learned and corrected for the impact of cryostat warmups, which are events that occurred approximately twice a year between 2015 and 2021 and introduce small jumps in measured radial velocities. These warmup dates are marked on the raw RVs in the top panel of Figure \ref{fig:detector warmups}, and on the predicted RVs from the CNN in the second panel. During periods of higher magnetic activity (the first half of the dataset), the RV offsets introduced by the warmups are smaller than the activity variations, so it is not easy to notice their effect. However, during periods of lower activity (the second half of the dataset), there are apparent jumps in the measured RVs occurring at the warm-up dates. This is illustrated by the two examples in the bottom two panels of Figure \ref{fig:detector warmups}. (A similar effect is also visible in Figure \ref{fig:low activity} at time 6840 days, or Julian Date 2458840). These two figures also show the neural network's predicted RVs at the times of the cryostat warmups which also show jumps, indicating that the network learned to predict instrumental artefacts in addition to the stellar activity signal.

The fact that the neural network can predict both instrumental and activity signals is interesting and suggests avenues for future improvement. The neural network was designed primarily to model stellar variability, and we therefore did not include features that directly trace the instrument's performance. As a result, we did not expect the neural network to mitigate instrumental-driven RV variations. Instead, it must have identified instrumental artifacts events based on shape changes in the white light CCFs, since that is the only input into our best model that is derived from HARPS-N. In the future, we may be able to enhance the ability of the network to remove artifacts by inputting instrument diagnostics like measurements from temperature and pressure sensors to the network.  This potentially makes neural networks an even more powerful tool for post-processing RV measurements, since identifying and removing instrumental artifacts will become even more critical as we try to find increasingly small planets with the next generation of extreme precision spectrographs. 

\subsection{Limiting Precision}

In this work, we have demonstrated an improvement in radial velocity scatter on the Sun to about 90 \cms. If we could achieve similar improvements in RV measurements of other stars, it would considerably improve our ability to detect small planets with the RV method.  However, the 90 \cms scatter in the corrected RVs is still far larger than the 10 \cms\ amplitude of an Earth-like planet around a Sun-like star.

We therefore wish to understand how to improve the precision further and why it is being limited. The corrected periodogram in Figure \ref{fig:periodogram} shows that most of the quasi-periodic stellar signals at the stellar rotation period have been effectively removed by the CNN model predictions although some residual still remains near 1/2 the rotation period. Moreover, in simulated data with only magnetic activity signals, \citet{deBeurs2022} were able to use CNNs to achieve RV precision of a few \cms using this method. This implies that we are able to remove most of the contribution of magnetic activity signals either at the stellar rotation period or in the long-term activity cycle, so much of the remaining RV scatter must be from some other source. 

There are two possible factors dominating the remaining RV scatter in the HARPS-N data: (i) instrumental noise and (ii) supergranulation\footnote{We note that granulation should be effectively mitigated in our solar data because we average over an entire day of observations, but it remains a major challenge for observations of other stars at night.}.  In terms of instrumental noise, $\sigma_{\text{instrumental}}$, HARPS-N requires frequent calibrations to ensure accuracy of the generated wavelength solutions, which limit the precision of the RVs. \citet{Dumusque2021} finds that the wavelength solutions from the HARPS-N DRS we use for our data changes by around 49 \cms\ on day-to-day timescales. While we showed that our neural network can predict and correct at least some instrumental artefacts, we do not expect that most of our input features to the CNN are able to trace the day-to-day variations in the instrument wavelength solution that dominate the scatter found by \citet{Dumusque2021}\footnote{Perhaps in future work, we may explicitly include features to predict instrumental noise contributions such as temperature sensor information or wavelength calibration data.}. The second factor likely dominating our remaining RV scatter is supergranulation, $\sigma_{\text{supergran}}$, which occurs on inactive regions of the solar surface and would not be traced well by many of the activity indicators in our model\footnote{\citet{Palumbo2024} found spectrographs of ultra-high resolution with $R \gtrsim 190,000 $ are required to detect granulation and supergranulation signatures.}. Additionally, supergranulation may manifest in the CCFs as translational shifts \citep{Meunier2020}, to which our model is not sensitive due to some of our data pre-processing steps described in Section \ref{sec:ccfs}.

Supergranulation results in RV variations of a similar scale to instrumental noise, with estimates ranging from 68 \cms by \citet{AlMoulla2023} to 86 \cms\ by \citet{Lakeland2023}. Likewise, we do not expect that most of the supergranulation signal can be predicted given our inputs; it is not magnetic, so most of our additional features do not apply, and it should produce only small changes in the shape of the white light CCF. The overall contribution to RV signals from these two factors is given by,
\begin{equation}
    \sigma_{\text{tot}} \approx \sqrt{\sigma^2_{\text{instrumental}} + \sigma^2_{\text{supergran}} },
\end{equation}
which for $\sigma_{\text{instrumental}}=49$ \cms\ and $\sigma_{\text{supergran}}=68\ -\ 86$ \cms\ gives $\sigma_{\text{tot}} =84\ -\ 99$ \cms. This is similar to the 92 \cms\ residuals of our CNN model.

\subsection{Future work and prospects}\label{sec:future work and prospects}
One avenue for future work is to apply the lessons learned from this investigation of solar data to observations of other stars. Already, the neural network white light CCF-based method of \citet{deBeurs2022} has been modified to search for planetary signals around other stars \citep{2022Zhao,deBeurs2024}. Other stars have smaller datasets, so modifying our method requires a linear regression approach, where a white light CCF-based stellar activity model and Keplerian signals are simultaneously fitted. This approach has the advantage of not requiring the prior removal of stellar activity signals before performing a planet search and allowing us to immediately apply these methods to other stars without requiring the extensive training set necessary for a neural network. Thus far, these methods have been successfully applied to 5 systems and reduced their RV scatter \citep{2022Zhao,deBeurs2024}. In the future, these linear regression methods could incorporate some of the features that we found to contain complementary information to the white light CCFs in this work. For example, we could add the chromatic CCFs, a proxy for the unsigned magnetic flux for stars beyond the Sun \citep[e.g.][]{Lienhard2023}, and H$\alpha$ into this linear regression and test how much this helps our stellar activity models for stars spanning the HR diagram. However, since other stars have smaller datasets, we will need to investigate dimensionality reduction techniques if we wish to add all of these additional features. 

In the future, we are also interested in looking for new input features that can be used to predict granulation and supergranulation for the Sun. Although testing the applicability of these methods developed for the Sun to other stars is essential, developing new methods by focusing on solar observations can also give insight into new ways of mitigating variability, which can later be applied on other targets. In this way, the Sun is a great test-bed for trying out new features to see if they can help improve RV precision, and identifying new correlations to stellar variability. Since we find that the current noise floor for solar observations appears to be dominated by supergranulation \citep{Lakeland2023}, we plan to look for new correlations with these phenomena in the future by using extremely high-resolution spectra at higher resolutions than HARPS-N's \citep[see e.g.][]{Palumbo2024}, or other observables, such as SDO's Dopplergrams or UV continuum observations. Although many of the types of observations SDO provides for the Sun may not be readily available for other stars, they can still help probe the underlying physics, find new correlations, and potentially inform the design of future extreme precision spectrographs to incorporate our knowledge of stellar variability.

\section{Conclusion} \label{sec: Conclusion}
To detect Earth-like exoplanets orbiting the habitable zone of Sun-like stars, we must first detect and remove the larger stellar activity signals that hide the planetary signals. These stellar activity signals are difficult to remove due to their quasi-periodic nature. Previously, \citet{deBeurs2022} showed that a ML algorithm can effectively remove stellar activity from observations of the Sun using shape changes in the white light CCFs. Here, we have extended this work and tested how much additional input features to the ML model help the model's ability to identify and remove stellar activity from observed RVs. 

We performed an extensive suite of tests where we added new features to a model similar to that of \citet{deBeurs2022}. We identified which features decreased RV scatter the most compared to a model with only white light CCFs. We found that the activity indicators that consistently gave the lowest $\sigma_\mathrm{percentile}$ values included the unsigned magnetic flux, the TSI, the TSI derivative, the S-Index, the chromatic CCFs, H$\alpha$ EW, and the FWHM and contrast of the white light CCFs. Our best-performing model used the white light CCFs from HARPS-N Solar Telescope \citep{Dumusque2015}, and additionally the disc-averaged, unsigned, unpolarized magnetic flux from the HMI aboard the SDO \citep{Scherrer2012}, and the TSI derivative from SORCE and TSIS-1. This model reduces the scatter in the RV data from 147.1 \cms\ to 93.3 \cms\ across 6 years of observations.

We find that the remaining RV variability left after removing our best model's predictions is likely dominated by instrumental shifts and supergranulation. While current and future spectrographs are believed or expected to have instrumental stability at the 10 \cms\ level, supergranulation may prevent further improvements in RV precision without advancements in its removal. There is a strong need for further study of the Sun to identify pathways to mitigate stellar RV variability due to superganulation before we can detect the 10 \cms\ signals of Earth analogues around Sun-like stars.

\vspace{0.1in}
\section*{Acknowledgements}

N.M. would like to thank the support from the International Research Opportunities Programme and the Department of Physics at Imperial College London, the Turing Scheme, MIT International Science and Technology Initiatives (MISTI), and MIT Department of Physics. N.M. is funded by a Science and Technology Facilities Council (STFC) studentship (ST/Y509231/1).

Z.L.D. would like to thank the generous support of the MIT Presidential Fellowship, the MIT Collamore-Rogers Fellowship and to acknowledge that this material is based upon work supported by the National Science Foundation Graduate Research Fellowship under Grant No. 1745302. Z.L.D. and A.V. acknowledge support from the D.17 Extreme Precision Radial Velocity Foundation Science program under NASA grant 80NSSC22K0848. A.C.C acknowledges support from STFC consolidated grant number ST/V000861/1
and UKRI/ERC Synergy Grant EP/Z000181/1 (REVEAL). This project has received funding from the European Research Council (ERC) under the European Union’s Horizon 2020 research and innovation programme (grant agreement SCORE No 851555). This work has been carried out within the framework of the NCCR PlanetS supported by the Swiss National Science Foundation under grants 51NF40\_182901 and 51NF40\_205606. This project has received funding from the Swiss National Science Foundation under the grant SPECTRE (No 200021\_215200). B.S.L. is funded by a Science and Technology Facilities Council (STFC) studentship (ST/V506679/1). B.S.L and A.M acknowledge funding from a UKRI Future Leader Fellowship, grant number MR/X033244/1. A.M acknowledges funding from a UK Science and Technology Facilities Council (STFC) small grant ST/Y002334/1.

The HARPS-N project has been funded by the Prodex Program of the Swiss Space Office (SSO), the Harvard University Origins of Life Initiative (HUOLI), the Scottish Universities Physics Alliance (SUPA), the University of Geneva, the Smithsonian Astrophysical Observatory (SAO), the Italian National Astrophysical Institute (INAF), the University of St Andrews, Queen's University Belfast, and the University of Edinburgh

We used \texttt{ChatGPT 5.1} \citep{OpenAIChatGPT} to improve the concision, clarity, and wording of parts of the manuscript.

\textit{Software:} NumPy \citep{harris2020array},
            Matplotlib \citep{Hunter2007},
            TensorFlow \citep{tensorflow2015-whitepaper},
            Astropy \citep{astropy:2013, astropy:2018, astropy:2022}
            SciPy \citep{Virtanen2020},
            specutils \citep{SpecUtils}.

\appendix
\section{}
\begin{figure*}
    \centering
    \includegraphics[width=\textwidth]{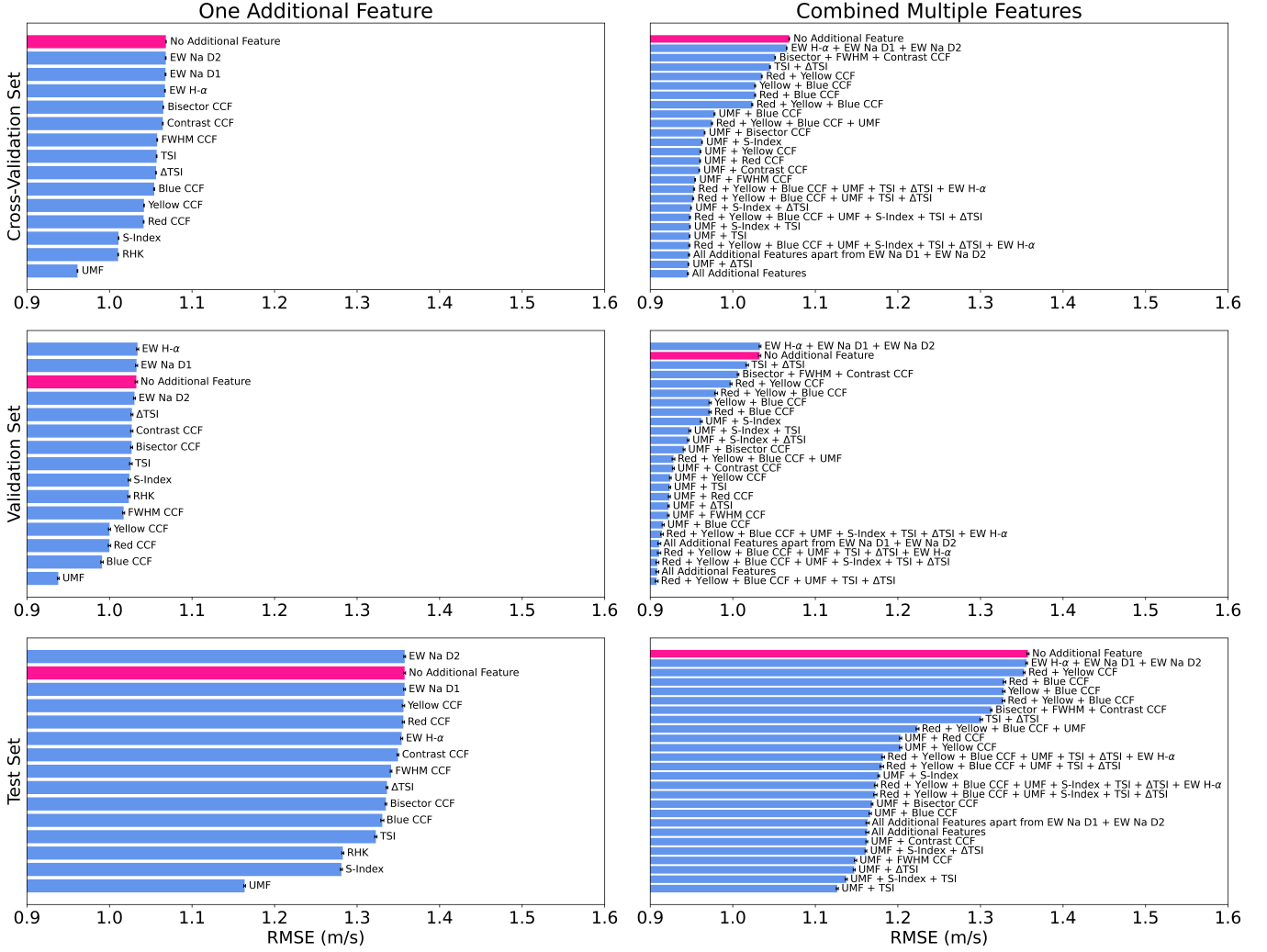}
    \caption{
    The median RMSE for the $\text{RV}_{\text{corrected}}$ values over $100$ runs for the cross-validation, validation, and test set for different model architectures (See Appendix \ref{sec: Appendix RMSE} for more details on the RMSE.). Models are shown with one additional feature to the original white light CCFs (left) and models with multiple features combined (right). The cross-validation set (top) contains 80\% of the data and is used to train the model. The validation set (middle) contains 10\% of the data and is used to evaluate the performance of the model on new inputs which can be used to optimize the model. The test set (bottom) has the remaining 10\% of the data, and is used to evaluate the final model performance on previously unseen data. A more detailed description of the three sets is provided in Section \ref{subsec: preparing training, validation, test sets}.
    The standard error of the median as given in Equation \ref{eq: standard error of median} are shown as black error bars.
    The model in pink uses only the original white light CCF's as the input to the CNN, and the models in blue represent those with any additional features.
    There are abbreviations for unsigned magnetic flux (UMF), Total Solar Irradiance (TSI), and TSI Derivative ($\Delta$TSI). 
    }
    \label{fig:bar chart results}
\end{figure*}

\begin{figure*}
    \centering
    \includegraphics[width=\textwidth]{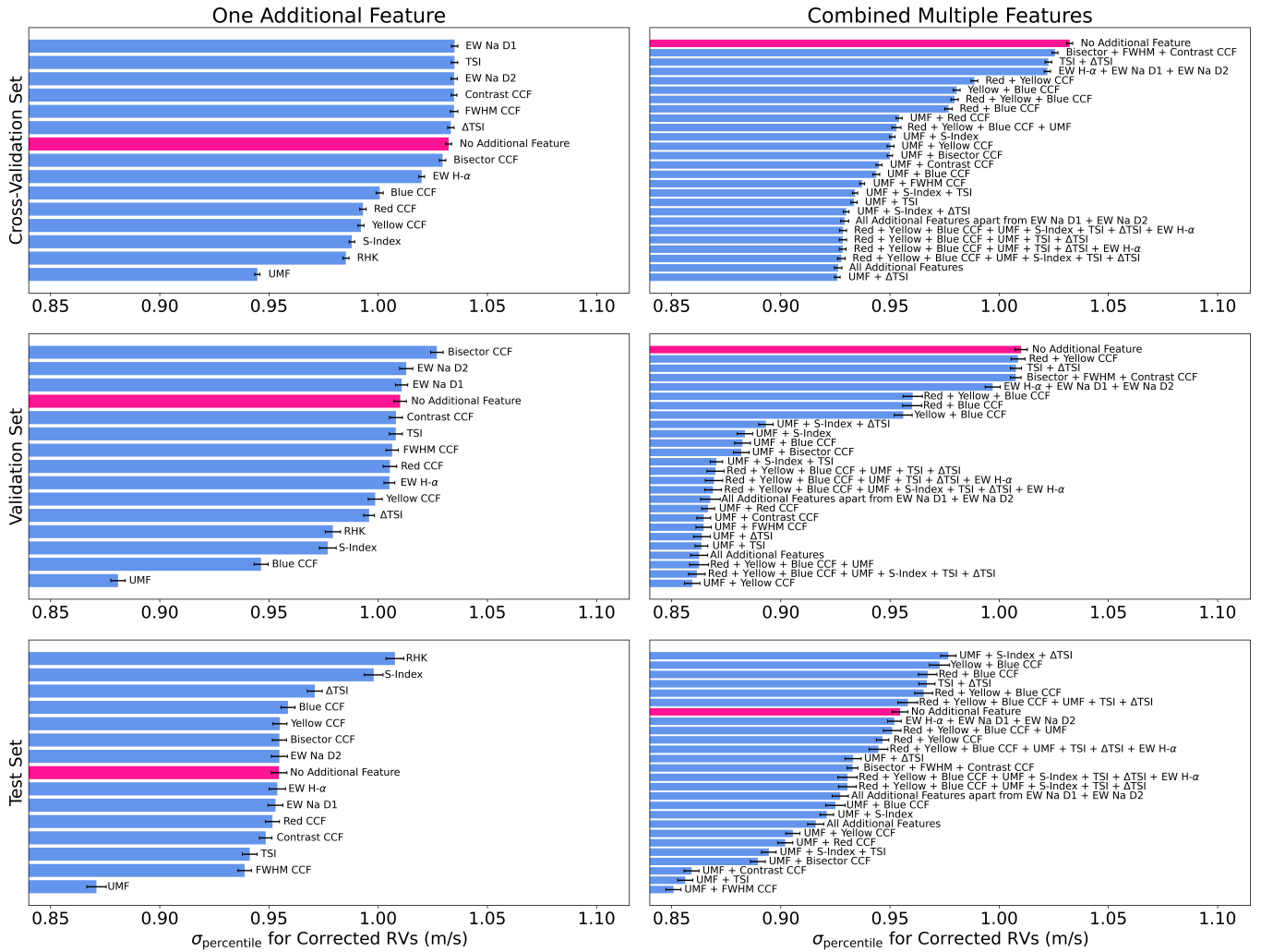}
    \caption{
    The same charts are shown as in Figure \ref{fig:bar chart results sigma percentile} using the same models, but showing the results of the median $\sigma_{\text{percentile}}$ values, where we use $\sigma_{\text{percentile}}$ as a less sensitive alternative to the standard deviation (see Appendix \ref{sec: Appendix sigma percentile} for more details). By comparing the models with the lowest $\sigma_{\text{percentile}}$ values, we can identify the most effective features for the NN. We find that the models containing UMF, TSI, $\Delta$TSI, EW H$\alpha$, and chromatic CCFs consistently result in the lowest $\sigma_{\text{percentile}}$ values.
    }
    \label{fig:bar chart results sigma percentile}
\end{figure*}

In this appendix, we include more detailed descriptions of the two performance metrics used in evaluating our models, the RMSE (Section \ref{sec: Appendix RMSE}) and $\sigma_\mathrm{percentile}$  (Section \ref{sec: Appendix sigma percentile}). In addition, we include the values of the performance metrics for each model in our analysis (Tables \ref{table:rmse values}, \ref{table:sigma percentile values} respectively) and figures to compare these values (Figure \ref{fig:bar chart results}, \ref{fig:bar chart results sigma percentile}).

\subsection{Root Mean Squared Error (RMSE)} \label{sec: Appendix RMSE}
The Root Mean Squared Error (RMSE) is a standard metric used to quantify the performance of a model by assessing the differences between the values predicted by the model and the values actually observed. The RMSE is defined as,
\begin{equation}
    \mathrm{RMSE} = \sqrt{\frac{1}{M}\sum_{i=1}^M(\mathrm{RV}{\mathrm{predicted},i} - \mathrm{RV}_{\mathrm{labels},i})^2},
\end{equation}
where $M$ represents the total number of samples being input or predicted. $\mathrm{RV}_{\mathrm{predicted},i}$ denotes the predicted value for the $i$th sample from the model, which in our analysis is a NN, which utilises input activity indicators for its predictions.
$\mathrm{RV}_{\mathrm{labels},i}$ denotes the actual, observed value for the $i$th sample, which for our NN models are the raw RVs without any activity corrections applied.

A lower RMSE value indicates a model that closely predicts the observed outcomes, reflecting its higher accuracy and performance. Therefore in our analysis, we conclude that the NN models with the lowest RMSE have the optimal model architecture or most effective input activity features.

\subsection{$\sigma_{\mathrm{percentile}}$} \label{sec: Appendix sigma percentile}
When selecting the optimal model architecture and identifying the most effective stellar activity features for our NN in predicting RVs, we initially rely on the RMSE metric as described above. However, RMSE is not resistant to the influence of outliers. We therefore also wish to consider outlier resistant metrics to assess how effective our models are.

The sigma percentile metric, $\sigma_\mathrm{percentile}$, is a robust statistical measure used to assess the variability of a dataset while minimising the influence of outliers. We define this as,
\begin{equation} \label{eq: sigma percentile}
    \sigma_\mathrm{percentile} = \frac{1}{2} ( \mathrm{RV}_\mathrm{84th\%} - \mathrm{RV}_\mathrm{16th\%} ) ,
\end{equation}
where $\mathrm{RV}_\mathrm{84th\%}$ and $\mathrm{RV}_\mathrm{16th\%}$ are the 84th and 16th percentiles of the RVs, respectively. This metric measures the spread of the middle 68\% of the data, providing a summary of data variability similar to the standard deviation in a normal distribution. However, unlike the standard deviation, $\sigma_\mathrm{percentile}$ is less sensitive to extreme values or outliers in the distribution, making it particularly useful for datasets with non-Gaussian characteristics.

\subsection{Standard Error of the Median} \label{sec: Appendix standard error on median}
When we determine the most effective stellar activity features, we find the $\sigma_\mathrm{percentile}$ of the corrected RVs 100 times for each model architecture. This is to ensure our results minimize the effect of random variations due to data ordering during training.
We then calculate the median and standard error of the median of the 100 $\sigma_\mathrm{percentile}$ values for each architecture.

The standard error of the median is a statistical measure used to quantify the uncertainty of the median estimate of a dataset. It is especially useful in situations where data distributions are not symmetric or contain outliers. It can be approximated for a dataset of $N$ samples by the formula:
\begin{equation} \label{eq: standard error of median}
\mathrm{Standard\,Error\,of\,Median} \approx 1.2533 \frac{\varsigma}{\sqrt{N}},
\end{equation}
where $\varsigma$ is the standard deviation of the dataset \citep{maindonald2010}.

\begin{table*}
\centering
\begin{tabular}{ccccc} 
\hline
      & \multicolumn{4}{c}{RMSE Median (\cms)}\\
      \hline
      Additional Input Features & Cross-Val Set & Validation Set & Test Set & Full Dataset \\
      \hline\hline
      Sodium D2 EW &  $106.82\pm0.04$ & $103.03\pm0.13$ & $135.77\pm0.11$ & $109.74\pm0.04$\\
      No additional features & $106.84\pm0.05$ & $103.26\pm0.15$ & $135.76\pm0.13$ & $109.74\pm0.04$\\ 
      Sodium D1 EW & $106.79\pm0.05$ & $103.28\pm0.13$ & $135.75\pm0.11$ & $109.65\pm0.04$\\
      H$\alpha$ EW & $106.73\pm0.05$ & $103.41\pm0.13$ & $135.40\pm0.13$ & $109.63\pm0.04$\\
      H$\alpha$ EW + Sodium D1 EW + Sodium D2 EW & $106.53\pm0.05$ & $103.30\pm0.11$ & $135.60\pm0.10$ & $109.47\pm0.04$\\
      CCF Contrast & $106.47\pm0.04$ & $102.70\pm0.14$ & $134.95\pm0.11$ & $109.26\pm0.03$\\
      CCF Bisector & $106.53\pm0.04$ & $102.68\pm0.12$ & $133.49\pm0.10$ & $109.17\pm0.03$\\
      CCF FWHM &  $105.78\pm0.05$ & $101.72\pm0.12$ & $134.12\pm0.11$ & $108.54\pm0.04$\\
      TSI Derivative & $105.64\pm0.04$ & $102.72\pm0.13$ & $133.63\pm0.11$ & $108.50\pm0.04$\\
      TSI & $105.71\pm0.04$ & $102.59\pm0.16$ & $132.27\pm0.14$ & $108.38\pm0.03$\\
      Blue CCF & $105.42\pm0.07$ & $99.10\pm0.16$ & $133.05\pm0.18$ & $107.93\pm0.06$\\
      CCF Contrast + CCF Bisector + CCF FWHM & $105.13\pm0.04$ & $100.64\pm0.01$ & $131.32\pm0.01$ & $107.59\pm0.03$\\
      Red CCF & $104.14\pm0.05$ & $100.00\pm0.15$ & $135.61\pm0.13$ & $107.33\pm0.04$\\
      Yellow CCF & $104.19\pm0.05$ & $100.01\pm0.13$ & $135.62\pm0.15$ & $107.32\pm0.04$\\
      TSI + TSI Derivative & $104.51\pm0.05$ & $101.74\pm0.16$ & $130.10\pm0.17$ & $107.04\pm0.04$\\
      Red CCF + Yellow CCF & $103.52\pm0.05$ & $99.78\pm0.12$ & $135.30\pm0.12$ & $106.76\pm0.04$\\
      Red CCF + Blue CCF & $102.71\pm0.06$ & $97.25\pm0.16$ & $132.93\pm0.17$ & $105.64\pm0.05$\\
      Yellow CCF + Blue CCF & $102.71\pm0.06$ & $97.26\pm0.15$ & $132.84\pm0.16$ & $105.62\pm0.06$\\
      Red CCF + Yellow CCF + Blue CCF & $102.36\pm0.07$ & $97.98\pm0.18$ & $132.79\pm0.17$ & $105.41\pm0.06$\\
      S-Index  &  $101.10\pm0.04$ & $102.41\pm0.15$ & $128.09\pm0.13$ & $104.25\pm0.04$\\
      Red CCF + Yellow CCF + Blue CCF + Unsigned magnetic flux & $97.48\pm0.07$ & $92.82\pm0.16$ & $122.38\pm0.18$ & $99.85\pm0.06$\\
      Unsigned magnetic flux + Blue CCF& $97.79\pm0.06$ & $91.57\pm0.13$ & $116.65\pm0.15$ & $99.22\pm0.05$\\
      Unsigned magnetic flux + S-Index & $96.28\pm0.04$ & $96.19\pm0.13$ & $117.68\pm0.11$ & $98.60\pm0.04$\\
      Unsigned magnetic flux + CCF Bisector & $96.59\pm0.05$ & $94.09\pm0.12$ & $116.88\pm0.11$ & $98.56\pm0.04$\\
      Unsigned magnetic flux + Yellow CCF & $96.08\pm0.06$ & $92.44\pm0.12$ & $120.33\pm0.15$ & $98.45\pm0.05$\\
      Unsigned magnetic flux + Red CCF & $96.04\pm0.05$ & $92.32\pm0.13$ & $120.33\pm0.13$ & $98.39\pm0.05$\\
      Unsigned magnetic flux & $96.12\pm0.04$ & $93.80\pm0.11$ & $116.38\pm0.11$ & $98.13\pm0.04$\\
      Unsigned magnetic flux + CCF Contrast & $95.94\pm0.04$ & $92.81\pm0.12$ & $116.24\pm0.11$ & $97.88\pm0.04$\\
      Red CCF + Yellow CCF + Blue CCF + Unsigned magnetic flux & $95.31\pm0.07$ & $91.08\pm0.20$ & $118.21\pm0.18$ & $97.44\pm0.06$\\
        + TSI + TSI Derivative + H$\alpha$ EW & \\
      Red CCF + Yellow CCF + Blue CCF + Unsigned magnetic flux & $95.20\pm0.08$ & $90.78\pm0.19$ & $118.05\pm0.21$ & $97.30\pm0.06$\\
        + TSI + TSI Derivative & \\
      Unsigned magnetic flux + CCF FWHM & $95.43\pm0.05$ & $92.19\pm0.10$ & $114.86\pm0.14$ & $97.22\pm0.04$\\
      Unsigned magnetic flux + S-Index + TSI Derivative & $94.94\pm0.04$ & $94.60\pm0.10$ & $116.13\pm0.11$ & $97.22\pm0.03$\\
      Unsigned magnetic flux + Red CCF + Yellow CCF + Blue CCF & $94.81\pm0.06$ & $90.86\pm0.19$ & $117.26\pm0.20$ & $96.95\pm0.06$\\
        + S-Index + TSI + TSI Derivative & \\
      Unsigned magnetic flux + S-Index + TSI & $94.81\pm0.04$ & $94.80\pm0.13$ & $113.75\pm0.14$ & $96.93\pm0.03$\\
      Red CCF + Yellow CCF + Blue CCF + Unsigned magnetic flux & $94.74\pm0.08$ & $91.42\pm0.17$ & $117.34\pm0.19$ & $96.86\pm0.07$\\
        + S-Index + TSI + TSI Derivative + H$\alpha$ EW & \\    
      Unsigned magnetic flux + S-Index + TSI + TSI Derivative   & $94.70\pm0.07$ & $91.10\pm0.19$ & $116.34\pm0.20$ & $96.71\pm0.06$\\
      + Red CCF + Blue CCF + Yellow CCF + H$\alpha$ EW\\
      + CCF Bisector + CCF FWHM + CCF Contrast \\
      Unsigned magnetic flux + S-Index + TSI + TSI Derivative & $94.56\pm0.07$ & $90.85\pm0.18$ & $116.30\pm0.19$ & $96.64\pm0.05$\\
       + Red CCF + Blue CCF + Yellow CCF \\
       + H$\alpha$ EW  + Sodium D1 EW + Sodium D2 EW \\
       + CCF Bisector + CCF FWHM + CCF Contrast \\
      Unsigned magnetic flux + TSI Derivative& $94.62\pm0.04$ & $92.22\pm0.10$ & $114.71\pm0.12$ & $96.60\pm0.04$\\
      Unsigned magnetic flux + TSI  & $94.77\pm0.04$ & $92.37\pm0.11$ & $112.66\pm0.13$ & $96.49\pm0.04$\\
      \hline
 \end{tabular}
 \caption{The median and error on the median of the RMSE values for the same models as in Table \ref{table:sigma percentile values} are shown sorted in order of decreasing median RMSE for the full dataset, with lower values indicating improved performance}.
 \label{table:rmse values}
 \end{table*}

\begin{table*}
\centering
\begin{tabular}{ccccc} 
      \hline
      & \multicolumn{4}{c}{$\sigma_{\text{percentile}}$ for Corrected RVs (\cms)}\\
      \hline
      Additional Input Features & Cross-Val Set & Validation Set & Test Set & Full Dataset\\
      \hline\hline
      Sodium D1 EW & $103.50\pm0.15$ & $101.07\pm0.28$ & $95.30\pm0.34$ & $102.97\pm0.12$\\
      Sodium D2 EW & $103.48\pm0.15$ & $101.29\pm0.30$ & $95.48\pm0.37$ & $102.84\pm0.14$\\
      TSI Derivative & $103.33\pm0.15$ & $99.58\pm0.25$ & $97.10\pm0.34$ & $102.71\pm0.13$\\
      CCF Contrast & $103.48\pm0.13$ & $100.81\pm0.30$ & $94.85\pm0.29$ & $102.71\pm0.12$\\
      No additional features & $103.23\pm0.14$ & $101.01\pm0.29$ & $95.46\pm0.36$ & $102.63\pm0.11$\\
      TSI & $103.50\pm0.15$ & $100.81\pm0.30$ & $94.12\pm0.34$ & $102.53\pm0.11$\\
      CCF FWHM & $103.48\pm0.18$ & $100.64\pm0.28$ & $93.89\pm0.31$ & $102.36\pm0.13$\\
      H$\alpha$ EW + Sodium D1 EW + Sodium D2 EW & $102.21\pm0.14$ & $99.71\pm0.35$ & $95.20\pm0.32$ & $102.13\pm0.12$\\
      TSI + TSI Derivative & $102.25\pm0.15$ & $100.77\pm0.25$ & $96.69\pm0.37$ & $102.11\pm0.12$\\
      H$\alpha$ EW & $101.99\pm0.13$ & $100.51\pm0.25$ & $95.38\pm0.37$ & $101.93\pm0.13$\\
      CCF Bisector & $102.95\pm0.15$ & $102.69\pm0.29$ & $95.48\pm0.32$ & $101.50\pm0.14$\\
      CCF Contrast + CCF Bisector + CCF FWHM & $102.56\pm0.14$ & $100.76\pm0.24$ & $93.29\pm0.25$ & $101.31\pm0.11$\\
      Red CCF & $99.30\pm0.15$ & $100.54\pm0.31$ & $95.16\pm0.32$ & $99.44\pm0.14$\\
      Yellow CCF & $99.21\pm0.14$ & $99.86\pm0.32$ & $95.50\pm0.32$ & $99.32\pm0.13$\\
      Blue CCF & $100.07\pm0.16$ & $94.64\pm0.33$ & $95.87\pm0.33$ & $99.21\pm0.15$\\
      Red CCF + Yellow CCF & $98.87\pm0.16$ & $100.86\pm0.32$ & $94.66\pm0.29$ & $98.73\pm0.14$\\
      S-Index & $98.80\pm0.13$ & $97.69\pm0.39$ & $99.80\pm0.43$ & $98.09\pm0.13$\\
      Red CCF + Yellow CCF + Blue CCF & $97.96\pm0.17$ & $96.04\pm0.44$ & $96.54\pm0.41$ & $97.60\pm0.16$\\
      Yellow CCF + Blue CCF & $98.05\pm0.16$ & $95.61\pm0.41$ & $97.27\pm0.46$ & $97.54\pm0.14$\\
      Red CCF + Blue CCF & $97.68\pm0.17$ & $96.02\pm0.44$ & $96.73\pm0.43$ & $97.01\pm0.14$\\
      Red CCF + Yellow CCF + Blue CCF + Unsigned magnetic flux & $95.30\pm0.21$ & $86.26\pm0.43$ & $95.10\pm0.40$ & $94.14\pm0.16$\\
      Unsigned magnetic flux + S-Index & $95.11\pm0.13$ & $88.36\pm0.35$ & $92.11\pm0.31$ & $93.81\pm0.13$\\
      Unsigned magnetic flux + Yellow CCF & $95.03\pm0.16$ & $85.95\pm0.35$ & $90.55\pm0.32$ & $93.60\pm0.14$\\
      Unsigned magnetic flux + Red CCF & $95.42\pm0.14$ & $86.67\pm0.30$ & $90.21\pm0.34$ & $93.60\pm0.14$\\
      Unsigned magnetic flux + CCF Bisector & $94.99\pm0.12$ & $88.19\pm0.36$ & $88.95\pm0.34$ & $93.54\pm0.11$\\
      Unsigned magnetic flux + Blue CCF & $94.37\pm0.16$ & $88.25\pm0.37$ & $92.51\pm0.44$ & $92.83\pm0.13$\\
      Unsigned magnetic flux & $94.47\pm0.12$ & $88.09\pm0.33$ & $87.10\pm0.45$ & $92.67\pm0.11$\\
      Unsigned magnetic flux + CCF Contrast & $94.50\pm0.14$ & $86.48\pm0.32$ & $85.91\pm0.34$ & $92.66\pm0.11$\\
      Unsigned magnetic flux + CCF FWHM & $93.72\pm0.12$ & $86.47\pm0.35$ & $85.09\pm0.35$ & $92.41\pm0.11$\\
      Unsigned magnetic flux + S-Index + TSI & $92.30\pm0.12$ & $87.05\pm0.29$ & $89.45\pm0.34$ & $92.54\pm0.10$\\
      Red CCF + Yellow CCF + Blue CCF + Unsigned magnetic flux & $92.84\pm0.16$ & $86.94\pm0.40$ & $94.47\pm0.43$ & $92.53\pm0.15$\\
        + TSI + TSI Derivative + H$\alpha$ EW \\
      Red CCF + Yellow CCF + Blue CCF + Unsigned magnetic flux & $92.84\pm0.17$ & $87.01\pm0.39$ & $95.82\pm0.46$ & $92.35\pm0.16$\\
        + TSI + TSI Derivative \\
      Unsigned magnetic flux + S-Index + TSI + TSI Derivative  & $92.93\pm0.18$ & $86.78\pm0.45$ & $92.72\pm0.37$ & $92.28\pm0.16$\\
      + Red CCF + Blue CCF + Yellow CCF + H$\alpha$ EW \\
      + CCF Bisector + CCF FWHM + CCF Contrast \\
      Unsigned magnetic flux + S-Index + TSI Derivative & $93.00\pm0.13$ & $89.32\pm0.33$ & $97.67\pm0.35$ & $92.23\pm0.11$\\
      Red CCF + Yellow CCF + Blue CCF + Unsigned magnetic flux & $92.85\pm0.16$ & $86.90\pm0.38$ & $93.06\pm0.45$ & $92.16\pm0.16$\\
        + S-Index + TSI + TSI Derivative + H$\alpha$ EW & \\
      Unsigned magnetic flux + TSI & $93.35\pm0.14$ & $86.37\pm0.29$ & $85.63\pm0.35$ & $92.10\pm0.12$\\
      Unsigned magnetic flux + S-Index + TSI + TSI Derivative & $92.63\pm0.17$ & $86.26\pm0.39$ & $91.60\pm0.37$ & $91.97\pm0.15$\\
      + Red CCF + Blue CCF + Yellow CCF \\
      + H$\alpha$ EW + Sodium D1 EW + Sodium D2 EW \\
      + CCF Bisector + CCF FWHM + CCF Contrast \\
      Unsigned magnetic flux + Red CCF + Yellow CCF + Blue CCF & $92.77\pm0.18$ & $86.16\pm0.38$ & $93.05\pm0.41$ & $91.92\pm0.16$\\
        + S-Index + TSI + TSI Derivative & \\
      Unsigned magnetic flux + TSI Derivative & $92.59\pm0.13$ & $86.39\pm0.38$ & $93.31\pm0.37$ & $91.57\pm0.11$\\

      \hline
 \end{tabular}
 \caption{The median and error on the $\sigma_{\text{percentile}}$ for the corrected RVs for $100$ models ran for each architecture are shown. They are in descending order of median $\sigma_{\text{percentile}}$ for the full dataset, with lower values indicating improved performance. Each architecture uses different additional input features. We find that the models containing unsigned magnetic flux, TSI, TSI Derivative, EW H$\alpha$, and chromatic CCFs consistently result in the lowest $\sigma_{\text{percentile}}$ values, and that the best model over the full dataset uses the unsigned magnetic flux and TSI derivative.}
 \label{table:sigma percentile values}
 \end{table*}

\bibliography{bibliography}{}
\bibliographystyle{aasjournal}

\end{document}